\documentclass[apj]{emulateapj}
\usepackage{graphics}
\usepackage{subfigure}

\shortauthors{Nevin et al.}
\shorttitle{The Origin of Double-Peaked Narrow Lines in AGN II: Kinematic Classifications for the Population at $z < 0.1$}

\begin{document}

\title{
 The Origin of Double-Peaked Narrow Lines in Active Galactic Nuclei II: Kinematic Classifications for the Population at $\lowercase{z} < 0.1$
}
\author{R. Nevin}
\affil{Department of Astrophysical and Planetary Sciences, University of Colorado, Boulder, CO 80309, USA}

\author{J. Comerford}
\affil{Department of Astrophysical and Planetary Sciences, University of Colorado, Boulder, CO 80309, USA}

\author{F. M{\"u}ller-S{\'a}nchez}
\affil{Department of Astrophysical and Planetary Sciences, University of Colorado, Boulder, CO 80309, USA}

\author{R. Barrows}
\affil{Department of Astrophysical and Planetary Sciences, University of Colorado, Boulder, CO 80309, USA}

\and

\author{M. Cooper}
\affil{Center for Cosmology, Department of Physics \& Astronomy, 4129 Reines Hall, University of California, Irvine, CA 92697, USA}

\begin{abstract}
We present optical longslit observations of the complete sample of 71 Type 2 active galactic nuclei (AGNs) with double-peaked narrow emission lines at $z < 0.1$ in the Sloan Digital Sky Survey. Double-peaked emission lines are produced by a variety of mechanisms including disk rotation, kpc-scale dual AGNs, and NLR kinematics (outflows or inflows). We develop a novel kinematic classification technique to determine the nature of these objects using longslit spectroscopy alone. We determine that 86\% of the double-peaked profiles are produced by moderate luminosity AGN outflows, 6\% are produced by rotation, and 8\% are ambiguous. While we are unable to directly identify dual AGNs with longslit data alone, we explore their potential kinematic classifications with this method. We also find a positive correlation between the narrow-line region (NLR) size and luminosity of the AGN NLRs (R$_{\mathrm{NLR}}\propto \; {\mathrm{L}_{\mathrm{[OIII]}}}^{0.21 \pm 0.05}$), indicating a clumpy two-zone ionization model for the NLR.
\end{abstract}

\keywords{galaxies: active -- galaxies: interactions -- galaxies: kinematics and dynamics -- galaxies: nuclei}

\section{Introduction}
A primary goal of modern astrophysics is to investigate how galaxies and their supermassive black holes (SMBHs) grow and coevolve. Correlations between the properties of the SMBH and the host galaxy suggest that the growth of galaxies and SMBHs are closely connected; for example, the M-$\sigma_{\star}$ relation connects the mass of the SMBH to the velocity dispersion of the stars in the galactic bulge (e.g., \citealt{Merritt2000, McConell2013}). Active galactic nucleus (AGN) feedback and AGN feeding processes have been invoked by theory as possible methods to maintain this relationship between SMBH and the host galaxy (e.g., \citealt{Croton2006,diMatteo2005,Springel2005}). 

`Positive' feedback, where the energy from the central AGN ignites star formation, has reproduced some observed relationships between the central AGN and the host galaxy (e.g., \citealt{King2005,Ishibashi2012,Silk2013}). The relative importance of positive or negative AGN feedback on the host galaxy remains unknown, as does the relative importance of different potential mechanisms for feedback (radiation, jets, or winds). 

AGN-driven `negative' feedback (henceforth, feedback) provides a method to evacuate gas from a galaxy and regulate star formation and the growth of the SMBH (e.g., \citealt{Croton2006,Springel2005,Hopkins2005}). The bimodal color distribution of galaxies in the nearby universe and the lack of massive galaxies in the mass function of galaxies require quenching of massive galaxies via a feedback mechanism (e.g., \citealt{Silk2011,Faber2007,Bell2004,Brown2007}). AGN feedback operates through a variety of mechanisms including relativistic plasma jets (e.g., \citealt{Fabian2012}), direct radiation (e.g., \citealt{Ciotti2010}), and mass outflows of ionized gas (e.g., \citealt{CK2003}). These types of feedback can impact material on different size scales from regions directly surrounding the central SMBH (\citealt{Tombesi2013}) to the Mpc-scale intergalactic medium (\citealt{Malarecki2013}). However, it is difficult to observe these different types of  feedback in action. Additionally, the details of the efficiency of energy and momentum transfer to the interstellar medium (ISM) and how this feedback might affect the surrounding galaxy are largely unknown (\citealt{Khalatyan2008,Hopkins2005}). Observational studies often focus on relativistic AGN jets (e.g., \citealt{Veilleux2005}) or other highly energetic cases such as Broad Absorption Line QSO outflows as examples of feedback (e.g., \citealt{Arav2013}). In this series of papers, we instead focus on the much more numerous (but less luminous) AGN-driven ionized outflows in the nearby universe. These types of galaxies account for 90-95\% of the total AGN population (\citealt{Rafter2009}), and their ionized mass outflows, or AGN winds, operate on a spatial scale coincident with circumnuclear star formation (\citealt{CF2015}).

In addition to feedback, AGN feeding through stochastic processes such as bar-driven inflows (e.g., \citealt{Coelho2011,Ellison2011}), or through merger-driven processes is important in SMBH-galaxy coevolution. Simulations of galaxy mergers indicate that during the merger, gas is driven to the center of the remnant galaxy (e.g., \citealt{Hopkins2005,Springel2005}). These simulations predict AGN feeding and SMBH growth will occur in this post-merger phase and observational studies have found that the AGN fraction increases from separations of 100 kpc to 10 kpc between stellar bulges in a galaxy merger (e.g., \citealt{Ellison2011,Koss2012,Ellison2013}). However, little is known about fueling at kpc-scale separations of these active nuclei, which we refer to as `dual AGNs'. 

Both feeding and feedback processes can be investigated using active galaxies that have a narrow line region (NLR) with disturbed kinematics. The NLR is a low density ($n_H \leq 10^6$ cm$^{-3}$) spatially extended region (from a few hundreds of parsecs to $\sim$ 30 kpc) surrounding the central AGN that is characterized by forbidden narrow emission lines ($\Delta v < 1000$ km s$^{-1}$, e.g., \citealt{Osterbrock2006,Schmitt2003,Hainline2013}). Disturbed kinematics manifest as double-peaked emission lines. These double-peaked narrow emission line profiles can be produced by dual AGNs, outflows, inflows, and disk rotation. Inflows and dual AGNs are associated with AGN feeding processes while outflows can probe AGN feedback.

In this paper we present a uniform sample of 71 double-peaked narrow emission line AGNs from the Sloan Digital Sky Survey (SDSS). Although this sample was originally selected with the purpose of identifying dual AGNs, this work is mostly concerned with the kinematics of single AGNs with disturbed NLRs. This sample of AGNs with disturbed NLR kinematics enables us to investigate the origin of the disturbed NLR and probe both feeding and feedback processes in these galaxies. 

However, determining the origin of double-peaked emission lines has been historically challenging since it is difficult to determine the spatial location of the emission in integrated spectra. For instance, ionized outflows have been identified based upon signatures in integrated spectra such as blue wings (e.g., \citealt{Whittle1985}). This is problematic since gas kinematics can shift across the spatial extent of the NLR, becoming impossible to disentangle in an integrated spectrum. As a result, past work has associated double-peaked emission lines with a variety of origins and often is unable to unambiguously identify the origin of these double peaks. Double-peaked and asymmetric emission lines have long been associated with an outflowing biconical structure; the blueshifted component is identified as the front-facing wall of the bicone, and the redshifted peak of the double-peaked profile is the rear-facing wall (e.g., \citealt{Heckman1981}, \citealt{Das2006}, \citealt{CF2015}). Other work has also suggested that double-peaked NLR emission lines could also be associated with kinematic dual AGNs (e.g., \citealt{Comerford2009a,Comerford2013,Barrows2013}). \citet{Smith2012} suggest that equal flux double-peaked emission lines could be associated with rotating disks.

Using spatially-resolved spectra we can now study the double-peaked profiles at each spatial position. In this work we develop a longslit kinematic classification technique to determine if the NLR kinematics are outflow-dominated or rotation-dominated. Here we focus on the kinematic nature as well as the ionization structure of the NLR. We use our observations of moderate luminosity AGNs to study the size-luminosity relationship for the NLR. Although the exact ionization structure of the NLR is not well determined, there is an observed positive correlation between the size of the NLR and the luminosity of the AGN, which indicates that the NLR is photoionized by the central AGN (e.g., \citealt{Hainline2014,Hainline2013,Bennert2002,Schmitt2003,Liu2013a,MS2015}). The slope of this relationship reveals the ionization structure of the NLR; a steeper slope of $R_{\mathrm{NLR}}\propto L_{\mathrm{[OIII]}}^{0.5}$ corresponds to a simplistic NLR described by a constant density law and a constant ionization parameter, while a shallower slope of 0.34 corresponds to a two-zone clumpier ionization model (\citealt{Baskin2005,Dopita2002}).

In this paper (Part 2 of a multiple paper series) we address the kinematic nature of the NLR for galaxies in the nearby universe for our sample of 71 Type 2 AGNs with double-peaked narrow emission lines at $z < 0.1$. We describe the sample selection and data reduction in Section~\ref{methods}. We describe our kinematic classification technique in Section~\ref{kin class}, where we classify galaxies as different subclasses of outflow- or rotation-dominated. We discuss our results from the kinematic classification and the properties of each kinematic class of galaxies in Section~\ref{all results}. In Section~\ref{all discussion}, we discuss the implications of the classification method for identifying dual AGNs as well as our measurement of the size-luminosity relation and its implications for the ionization structure of the NLR. We present our conclusions in Section~\ref{conclude}. A cosmology with $\Omega_m = 0.3$, $\Omega_{\Lambda}=0.7$, and $h=0.7$ is assumed throughout.

\section{Methods}
\label{methods}

\subsection{Sample Selection}
The 71 galaxies in the uniform sample originate from a full sample of Type 2 AGNs with double-peaked [OIII] emission lines in SDSS (\citealt{York2000}). Three groups selected catalogues of double-peaked AGNs (\citealt{Wang2009,Liu2010,Smith2010}).

\begin{deluxetable*}{llll}
\tabletypesize{\scriptsize}
\tablewidth{0pt}
\tablecolumns{4}
\tablecaption{Summary of Longslit Observations}
\tablehead{
\colhead{SDSS ID} & \colhead{Galaxy Name} & \colhead{$z$} & \colhead{Telescope/Instrument}}
\startdata
J000249.07+004504.8 & J0002+0045 & $0.086735 \pm 3.9\mathrm{E}-5$& Lick/Kast \\
J000911.58$-$003654.7 & J0009$-$0036 & $0.073135 \pm 3.4\mathrm{E}-5$& MMT/Blue Channel  \\
J013555.82+143529.7 & J0135+1435 & $ 0.072157 \pm 1.9\mathrm{E}-5$ &Lick/Kast  \\
J015605.14$-$000721.7 & J0156$-$0007 & $ 0.080964 \pm 1.8\mathrm{E}-5$ &Lick/Kast  \\

\enddata
\tablecomments{Optical longslit observations of double-peaked AGNs.
Column 1: SDSS ID, also RA and Dec.
Column 2: Redshifts and errors, determined from the velocity of the stellar absorption lines.
Column 3: Observatory and spectrograph
(This table is available in its entirety in the online journal.)
} 
\label{obs} 
\end{deluxetable*}

\citet{Wang2009} selected 87 Type 2 active galaxies using BPT emission-line diagnostics (\citealt{Baldwin1981}), then made a cut to eliminate galaxies with SDSS $r$-band magnitude $r > 17.7$. They selected for similar intensity peaks of the double-peaked profiles using a flux ratio cut of 1:10 between the intensity of each peak and required a wavelength separation between these two peaks of $\Delta \lambda \geq 1$ \AA. \citet{Smith2010} selected a sample visually for active galaxies that exhibit double peaks. However, the Type 1 and Type 2 AGNs from \citet{Smith2010} are located at redshifts $0.1 < z < 0.7$, so they are not included in the sample selection for this paper. Since \citet{Smith2010} is the only catalog with Type 1 objects, by excluding these higher redshift objects we also restrict our sample to Type 2 objects. This avoids the influence of broad lines ($\Delta v > 1000$ km s$^{-1}$) on the [OIII] profiles. \citet{Liu2010} selected 167 Type 2 AGNs by making a S/N $>$ 5 cut for [OIII]$\lambda 5007$ and requiring that both [OIII]$\lambda$5007 and [OIII]$\lambda$4959 be best fit by two Gaussians. This excludes AGNs with more complex profiles, wings, and $>2$ Gaussian components. These three groups selected 340 unique objects. We select the complete sample of 71 double-peaked Type 2 AGNs that are at $z < 0.1$ to ensure sub-kpc spatial resolution on all optical longslit instruments used in the follow-up observations.

To determine the nature of these objects and further characterize their properties, we observe them using two complementary follow-up methods: optical longslit spectroscopy and Jansky Very Large Array (VLA) radio observations.\footnote{Spatially-resolved imaging of X-ray sources (e.g., \citealt{Komossa2003,Mazzarella2012,Barrows2016}) that are coincident with emission-line peaks can also confirm the presence of dual AGNs. In this paper series we focus on radio observations as a method for confirming dual AGNs.} We use the longslit spectroscopy to map the source of the double-peaked emission across the spatial extent of each galaxy; to do so we observe each galaxy at two position angles in order to constrain the orientation and spatial positioning of the NLR. For most galaxies, these two position angles are orthogonal with the exception of galaxies that have an intriguing disturbed feature or companion galaxy at a given position angle. 

We choose one position angle to be the photometric major axis of the galaxy from the SDSS $r$-band photometry. This is motivated by one of our science goals, which is to determine if the NLR is rotational in origin. Galaxies in which the NLR is dominated by rotation will demonstrate the most extended emission along the photometric major axis, in the plane of the galaxy.

We use various spectrographs with similar pixelscales (Lick Kast Spectrograph, 0.78$^{\prime \prime}$/pixel; Palomar Double Spectrograph, 0.39$^{\prime \prime}$/pixel (\citealt{Oke1982}); MMT Blue Channel Spectrograph, 0.29$^{\prime \prime}$/pixel (\citealt{Schmidt1989}); APO Dual Imaging Spectrograph, 0.42$^{\prime \prime}$/pixel in the blue channel, 0.4$^{\prime \prime}$/pixel in the red channel; and Keck DEep Imaging Multi-Object Spectrograph, 0.12$^{\prime \prime}$/pixel (\citealt{Faber2003}). We use a 1200 lines mm$^{-1}$ grating for all spectrographs. Table~\ref{obs} lists the details of these observations.

The VLA observations are complementary to the optical longslit data and together they can fully constrain the source of the double peaks for a given galaxy; \citet{MS2015} focus on this technique for a subsample of 18 double-peaked AGNs. In this paper we develop a novel technique using only the longslit data to characterize the NLR of this sample of 71 galaxies. In future work, we will combine our longslit observations with the VLA data for the full sample of 71 galaxies (M{\"u}ller-S{\'a}nchez et al. in prep.).

\subsection{Data Reduction}
We reduce and extract both the 2 and 1-dimensional spectra at both position angles for each galaxy using the IRAF packages CCDPROC and APALL, respectively. The Keck/DEIMOS data were reduced using the DEEP2 pipeline (\citealt{Cooper2012,Newman2013}). We preserve the spatial information in the 2d spectra and use the aperture-extracted 1d spectra for wavelength solutions, which we use to produce the velocity maps. We obtain accurate systemic velocities using extinction corrected stellar absorption features from the OSSY SDSS DR7 value-added catalog (\citealt{Oh2011,Abazajian2009}) and the IDL code GANDALF (\citealt{Sarzi2006}).

\subsection{Extracting and Characterizing [OIII] Profiles}
\label{kin setup}

\begin{figure*}
\begin{center}
\includegraphics[scale=0.30]{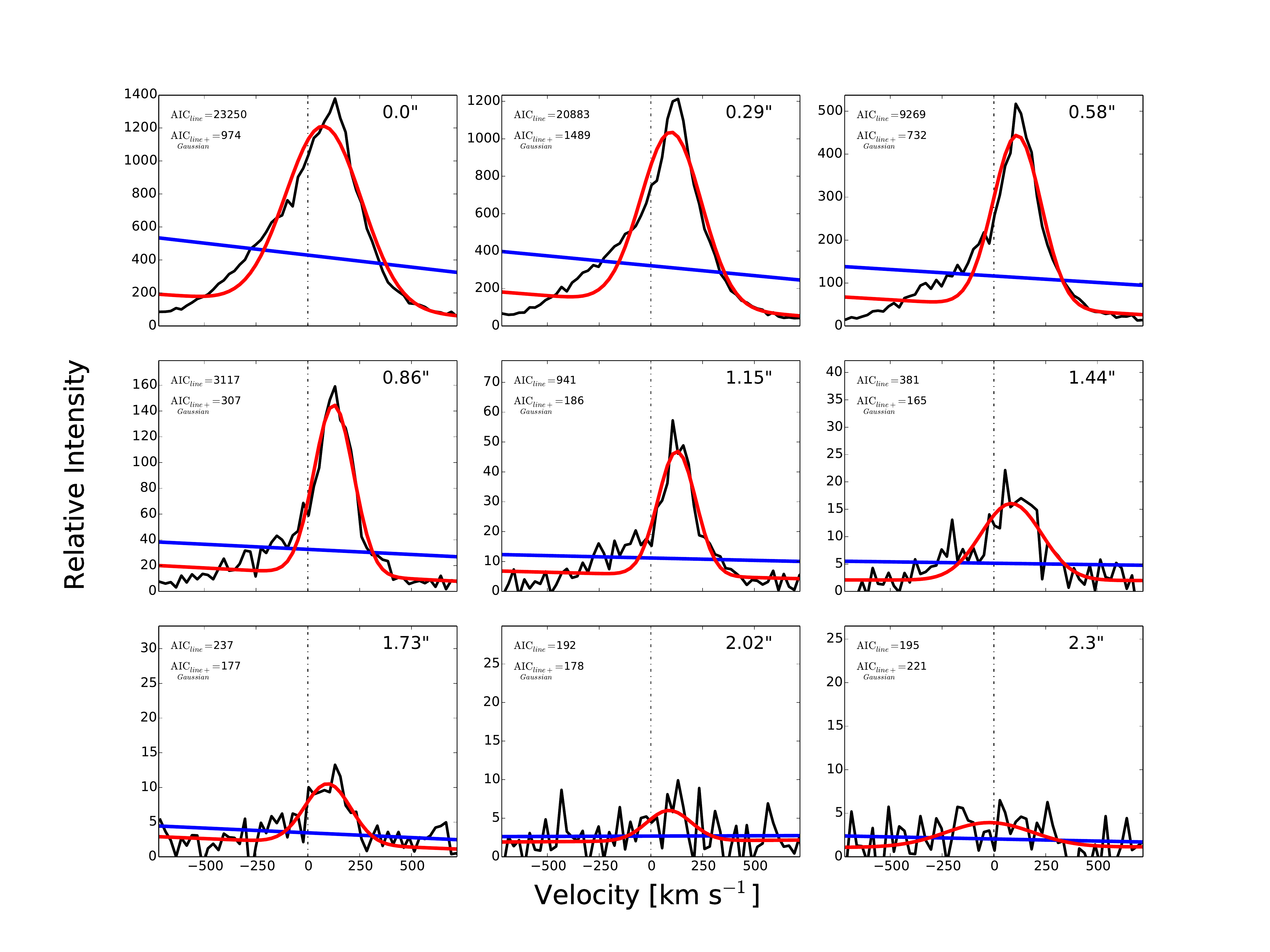}
	
\caption{The Akaike statistics demonstrated for the [OIII]$\lambda$5007 emission for the example galaxy J0009-0036. For each spatial row of the example galaxy J0009-0036 (to the NE of the spatial center of the galaxy) we report the spatial position (in arcseconds, upper right). Each panel is spaced by angular distance 0.29$^{\prime \prime}$, which is the pixelscale of the MMT Blue Channel Spectrograph. We start the figure at spatial position 0.0$^{\prime \prime}$ to highlight the diminishing flux of the emission line to one spatial extreme (the NE) of the galaxy spatial center. Note that this is a symmetric NLR with flux also extending to the SW of the spatial center. We show the Akaike statistic for both a line (two parameter fit), plotted in blue, and a single Gaussian with an underlying line (five parameter fit), plotted in red. If $\mathrm{AIC}_{line+Gaussian} < \mathrm{AIC}_{line}$, then that spatial row is considered to have significant emission and lies within the Akaike width. Here, the 2.02$^{\prime \prime}$ row is the row of last significant emission, and the 2.3$^{\prime \prime}$ row no longer has significant emission. We repeat this process for both observed position angles of each galaxy. \label{fig:extent}  }
\end{center}

\end{figure*}

With fully reduced data in hand, we use a variety of IDL and Python programs to fit the [OIII]$\lambda 5007$ line profiles at each spatial position along the slit and extract velocity and dispersion information. We determine the spatial center of emission for each 2d spectrum using the stellar continuum. We fit a one dimensional Gaussian across the continuum in two 10 pixel cutouts on either side of the wavelength center of the [OIII]$\lambda5007$ profile, and determine its average center and full width at half maximum (FWHM$_{\mathrm{cont}}$). We later apply this spatial center and width (FWHM$_{\mathrm{cont}} \pm$1 row) when we refer to the `resolved center' of the emission. The positive spatial direction shown in all plots of the 2d spectra in this work corresponds to the NE direction on the sky. 

We use an information criterion to determine both the extent of the emission and the number of Gaussian components to fit at each spatial position. The Akaike Information Criterion (AIC) is a least squares statistic that introduces a penalty for additional parameters, defined by \citet{Akaike1974}. Numerically, $\mathrm{AIC} = \chi^2 + 2k$, where $k$ is the number of parameters and $\chi^2$ is the chi-square statistic:

$$\chi^2=\sum_{i=1}^n \frac{(y_i-f(x_i;\hat{\theta}))^2}{\hat{\sigma}_i^2}$$where $n$ is the number of data points in the sample, $y_i$ is the measured flux, $f(x_i; \hat{\theta})$ is the Gaussian model for the emission line flux, $x_i$ is the input wavelength of the Gaussian model,  $\hat{\theta}$ are the parameter values of the Gaussians, and $\hat{\sigma}_i^2$ is the measurement uncertainty on the measured flux. We utilize the corrected AIC ($\mathrm{AIC}_c$) due to our finite number of data points: $\mathrm{AIC}_c = \mathrm{AIC} + 2k(k+1) / (n-k-1)$ where $n$ is the sample size and $k$ is the number of parameters. When comparing two models with a different number of parameters, the model that produces the smallest value for the above statistic represents the better fit. 

\begin{deluxetable*}{lllllll}

\tabletypesize{\scriptsize}
\tablewidth{0pt}
\tablecolumns{7}
\tablecaption{Measured Luminosities}
\tablehead{
\colhead{Galaxy} & 
\colhead{L$_{\mathrm{[OIII]}}$ [erg s$^{-1}$] } &  
\colhead{L$^c_{\mathrm{[OIII]}}$ [erg s$^{-1}$]} & 
\colhead{L$_{\mathrm{bol}}$ [erg s$^{-1}$]} & 
\colhead{kpc/$^{\prime \prime}$ } &   
\colhead{Aik Width [pixels]} & 
\colhead{R [pc]}  }
\startdata 
J0002+0045 & $(2.252\pm 0.397) \times 10^{41}$ & $(1.516\pm 0.267) \times 10^{42} $& $(6.886\pm 1.216) \times 10^{44} $& 1.62 & $11 \pm 2 $ & $6330 \pm 980$\\
J0009$-$0036 & $(4.813\pm 0.420) \times 10^{41}$ & $(4.444\pm 0.388) \times 10^{42} $& $(2.0178\pm 0.1764) \times 10^{45}$ & 1.39 & $19 \pm 2 $ & $3610 \pm 410$\\
J0135+1435 & $(1.104\pm 0.142) \times 10^{41}$ & $(4.52\pm 0.58) \times 10^{41} $& $(6.42\pm 0.82) \times 10^{43}$ & 1.37 & $12 \pm 1 $ & $6430 \pm 800$\\
J0156$-$0007 & $(1.420\pm 0.221) \times 10^{41}$ & $(5.69\pm 0.88) \times 10^{41} $& $(8.08\pm 1.26) \times 10^{43} $& 1.53 & $14 \pm 1 $ & $8330 \pm 810$\\

\enddata
\tablecomments{Measured properties for all galaxies. Column 1: Galaxy name. Column 2: Observed [OIII] luminosity before it is dereddened. Column 3: Dereddened [OIII] luminosity. Column 4: Bolometric luminosity. Column 5: Kpc per arcsecond conversion. Column 6: Akaike width of the NLR in pixels. Column 7: The corresponding radius of the NLR in parsecs. This is R$_{\mathrm{NLR}}$ in the size-luminosity relationship. (This table is available in its entirety in the online journal.)}
\label{R-L}
\end{deluxetable*}

\begin{figure*}
\begin{center}
\includegraphics[scale=0.35]{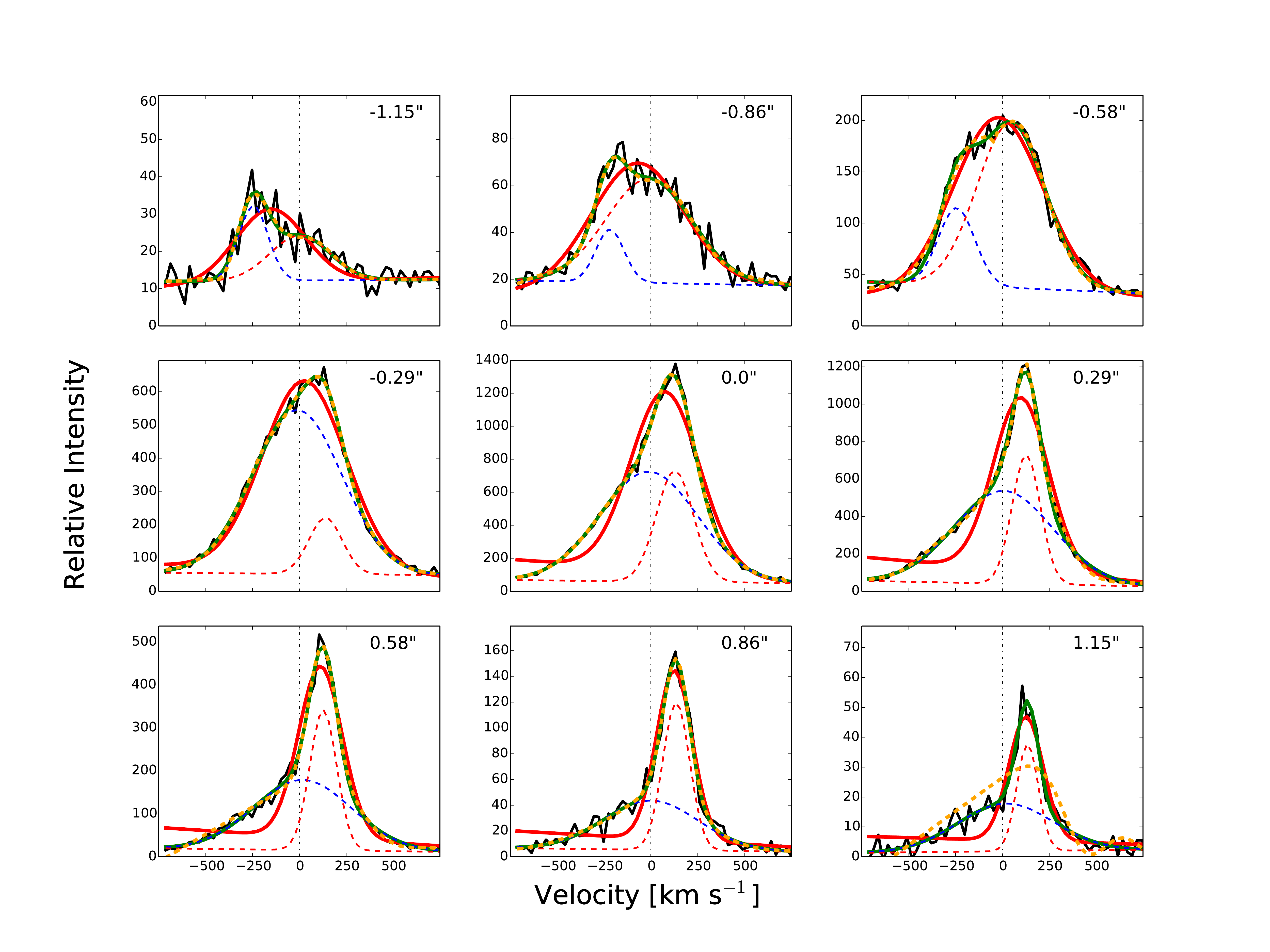}
\caption{Fitting multiple Gaussians across the spatial extent of J0009-0036 to the [OIII]$\lambda$5007 profile. The galaxy and position angle of the observation are the same as Figure~\ref{fig:extent} but now we change the spatial coverage to 1.15$^{\prime \prime}$ on either side of the spatial center of the galaxy to highlight the fitting of multiple Gaussian components. The data are shown as solid black lines. For each spatial row, we show the integrated one, two, and three Gaussian fits in red solid, green solid, and yellow dashed lines, respectively. We also display the individual Gaussians from the two Gaussian fit in blue and red dashed lines, where the blue represents the blueshifted component and the red is the redshifted component of the two Gaussian fit. The black dashed vertical line is the systemic velocity of the galaxy. Intensity is displayed in units of relative intensity. This plot demonstrates the $\mathrm{AIC}_c$ applied to determining the number of Gaussians to fit to each significant spatial row (`significant' implies that the emission line is already better fit by one than zero Gaussians and is therefore within the Akaike width, Figure~\ref{fig:extent}).\label{Gaussians} }
\end{center}

\end{figure*}

We apply this statistic to each row in the spatial direction to determine the extent of the emission. We establish if the row is better fit with a two parameter linear fit or a five parameter Gaussian + linear fit (Figure~\ref{fig:extent}). This provides what we define as the `Akaike width' of the emission, which is measured in both arcseconds and pixels, and encompasses all the rows that are best fit by a five parameter Gaussian + linear fit. We then derive errors on this measurement from a Monte Carlo simulation. We construct 100 realizations of the spectrum by adding Gaussian noise with flux according to the inverse variance error image, and repeat the measurement of the Akaike width for each realization. The mean value of the Akaike width and its standard deviation are derived from the properties of the resulting distribution (Table~\ref{R-L}).

We also apply the Akaike statistic to determine the appropriate number of Gaussians to fit to each spatial row (Figure~\ref{Gaussians}). We use the $\mathrm{AIC}_c$ to determine how many Gaussians are the best fit for each row within the resolved center of the galaxy (FWHM$_{\mathrm{cont}} \pm$1 row). We classify a galaxy as having $> 2$ Gaussian components only if more than half of these interior rows are better fit by three or more components. This approach is used in the kinematic classification scheme.

\begin{figure}[Ht]
\begin{center}
\includegraphics[scale=0.4]{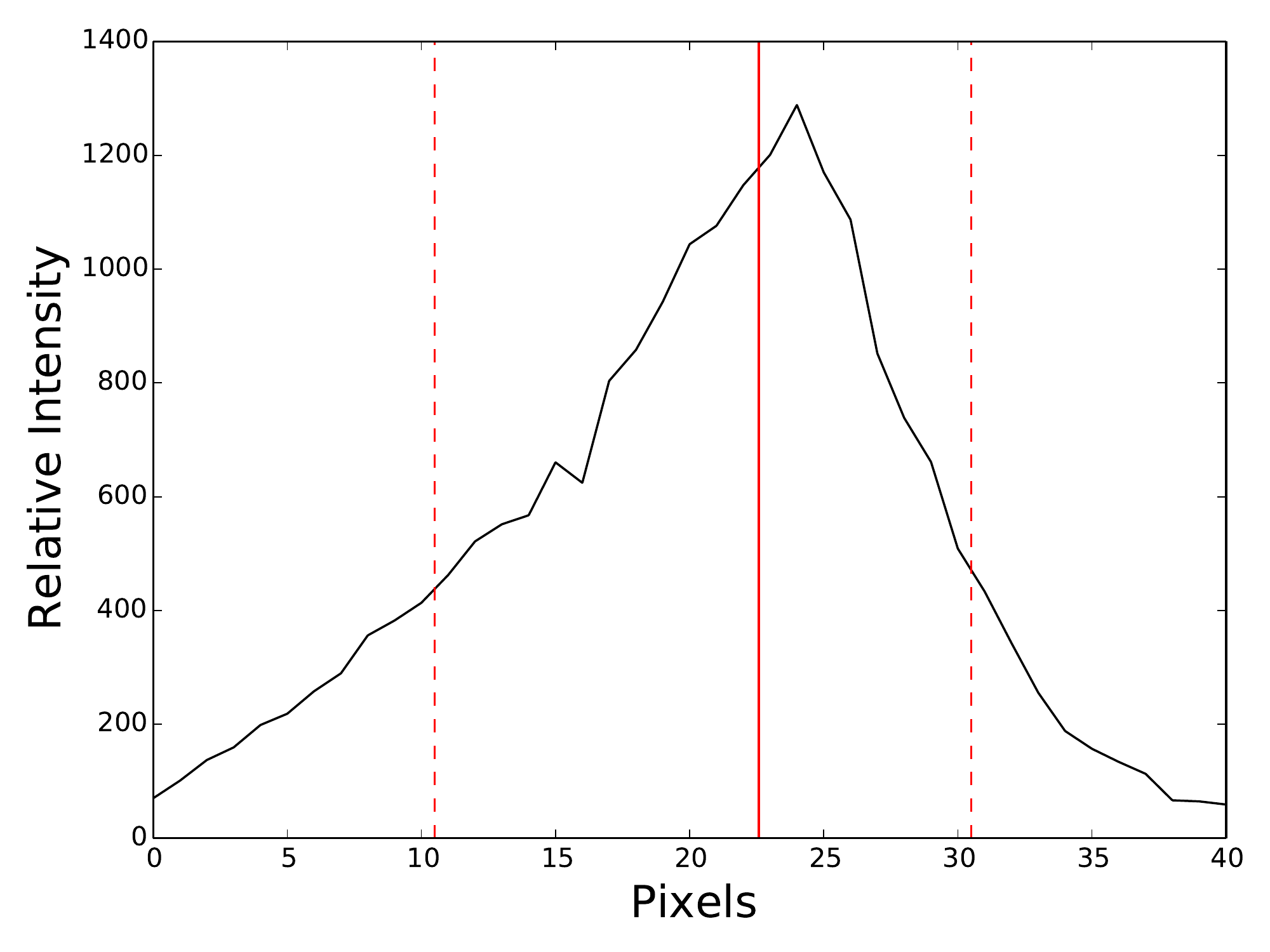}
\caption{ The asymmetry measurement reveals a blue wing for J0009-0036. Here, the [OIII]$\lambda 5007$ profile of the central spatial row is plotted in black for a 40 pixel cut (20 pixels on either side of the systemic velocity). The red solid line represents the median velocity ($v_{med}$) for the profile; this is the velocity that corresponds to a summed 50\% of the integrated flux profile at this spatial position. The left red dashed line is $v_{10}$, or the velocity that corresponds to 10\% of the integrated flux profile, and the right dashed line corresponds to $v_{90}$. For this galaxy, $A=-0.146$, indicating the presence of a blue wing.\label{assy} }
\end{center}
\end{figure}

 We extract velocity information for each row within the Akaike width by fitting both one and two Gaussians. The velocity offsets of each of the one and two Gaussian profiles are calculated relative to the systemic velocity of the galaxy. The center of each of these Gaussians in velocity space is mapped across the spatial extent of the galaxy. Similarly, we map the dispersion of each of these Gaussian profiles across the galaxy (reported as $\sigma$ for the one Gaussian fit and $\sigma_1$ and $\sigma_2$ for the two Gaussian fit). We derive errors on these calculations from the Monte Carlo method described above. The uncertainty on velocity measurements increases as the distance from the galaxy center increases because the S/N for these rows drops dramatically.

\begin{figure}[b]
\begin{center}
\includegraphics[scale=0.40]{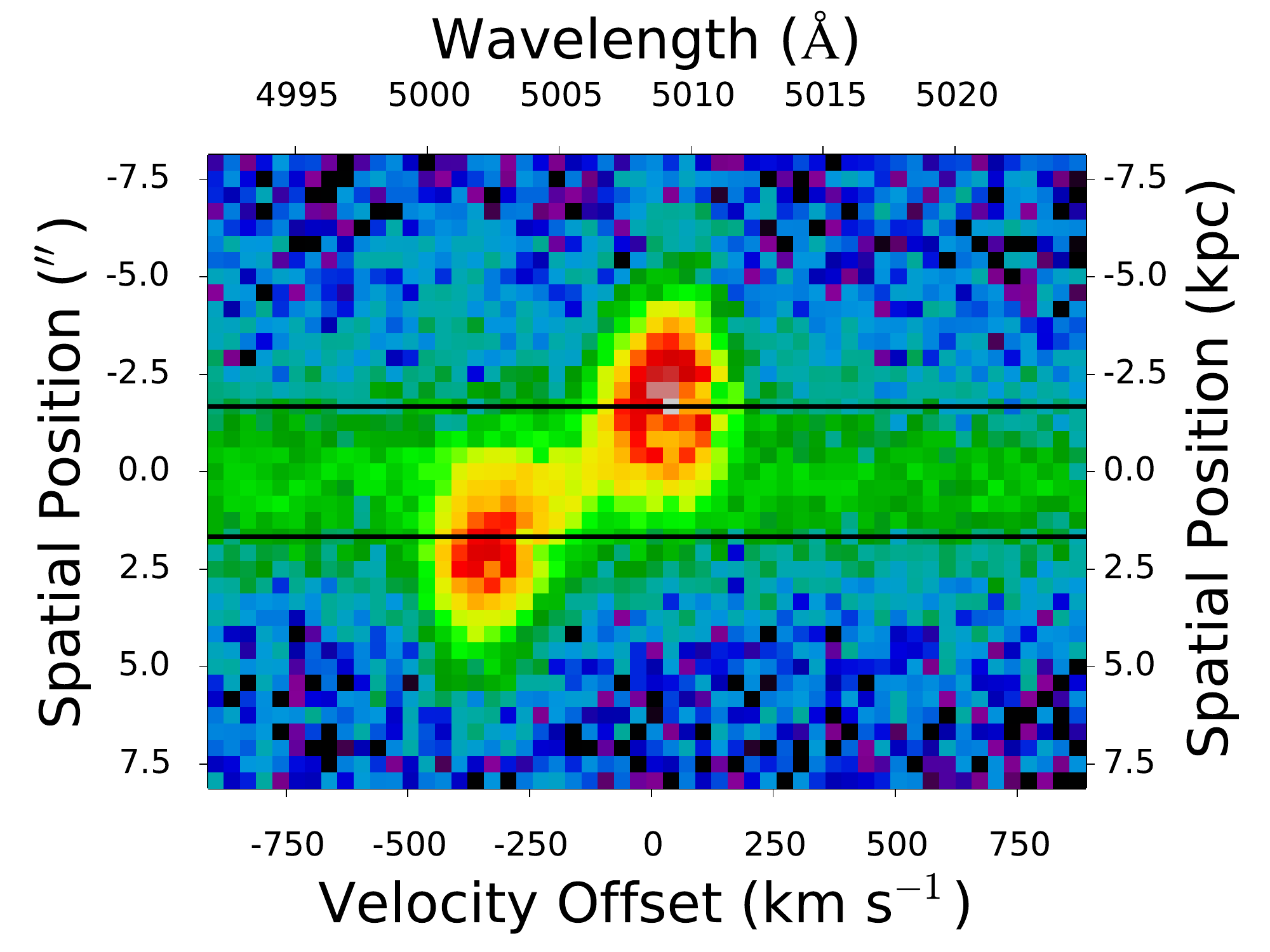}
\includegraphics[scale=0.40]{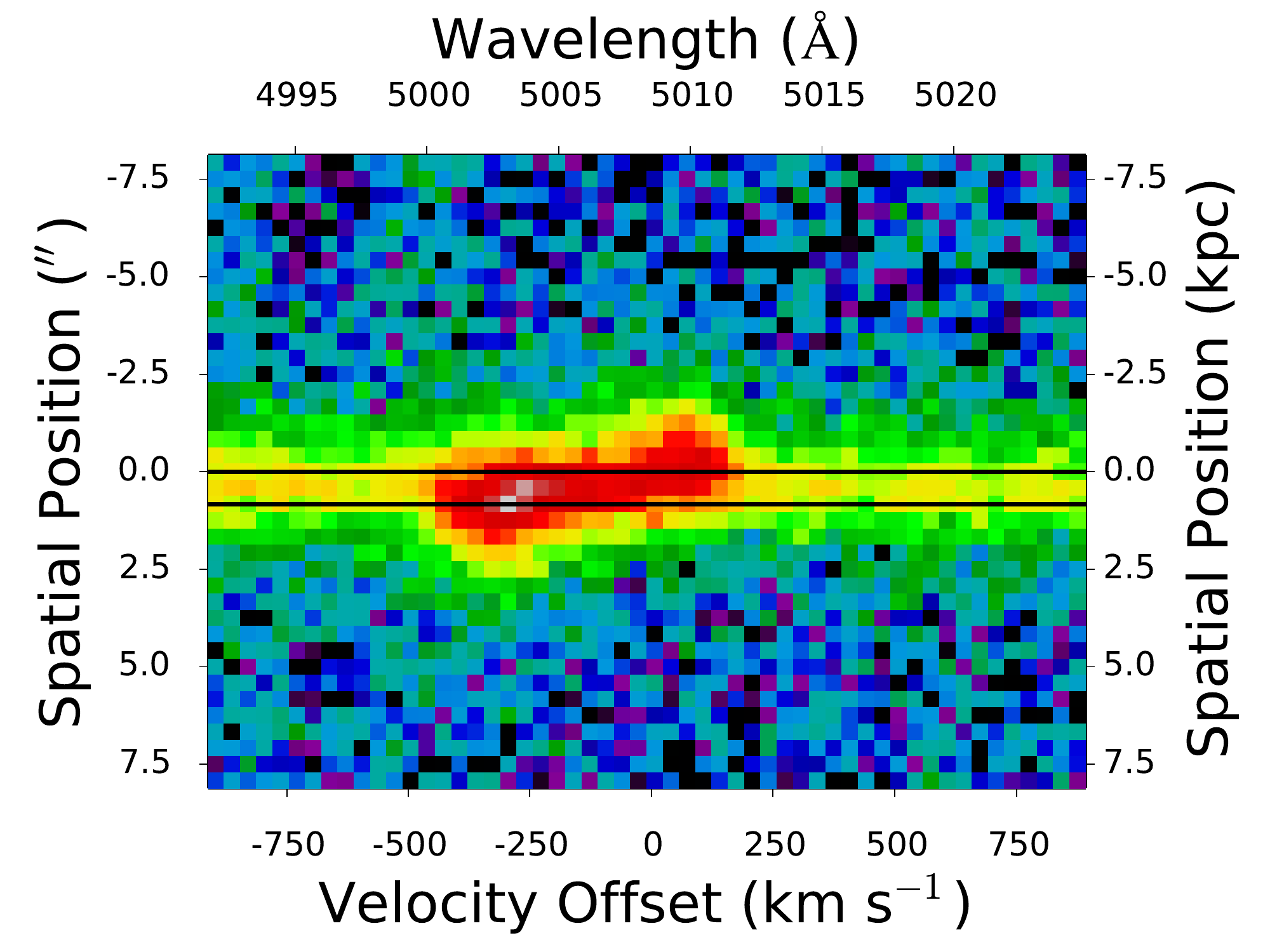}
	
\caption{Spatial centroiding technique for J1516+0517. [OIII]$\lambda 5007$ images centered vertically on the spatial center of J1516+0517 for both position angles (top: PA=81$^{\circ}$ East of North, and bottom: PA=171$^{\circ}$). The black lines show the fitted spatial position for each of the two spectral components ($\lambda >5007$ and $\lambda < 5007$ here). PA=81$^{\circ}$ displays a greater separation. Using both of these individual PAs, we are able to reconstruct the separation on the sky as well as the position angle on the sky of maximal separation of the double emission components. \label{a}}
\end{center}

\end{figure}

 Outflows and NLRs with disturbed kinematics have asymmetric emission line profiles and often blue wings. We utilize a nonparametric diagnostic for profile asymmetry. In order to avoid excluding profiles that have low flux wings or other non-traditional types of asymmetry, we choose not to assign parametric Gaussian metrics of asymmetry. Instead, we employ the nonparametric measurement for line profile asymmetry from \citet{Liu2013}: $$A\equiv \frac{(v_{90}-v_{med})-(v_{med}-v_{10})}{W_{80}}$$
where $v_{10}$ and $v_{90}$ are the velocities that encompass 10\% and 90\% of the integrated flux across the profile, respectively, $v_{med}$ is the velocity that corresponds to the median value of the integrated flux profile, and $W_{80}$ is defined as $W_{80}\equiv v_{90}-v_{10}$. The sign of asymmetry is negative if the profile has a blue wing and positive for red wings.

This statistic (Figure~\ref{assy}) is sensitive to double-peaked profiles that have unequal flux ratios; specifically the absolute value of asymmetry is large if the profile has what can be better described as a shoulder as opposed to an equal flux double-peaked profile. We only use the measurement to classify rotation-dominated profiles as disturbed or obscured. We discuss this aspect of kinematic classification in Section~\ref{kin class}.

In addition, we measure the position angle of the NLR [OIII]$\lambda 5007$ emission (PA$_{\mathrm{[OIII]}}$) on the sky. This allows us to determine if the [OIII] emission is rotational in origin. We fit position centroids at each peak of the double-peaked profile and calculate a separation in spatial position (Figure~\ref{a}). By iterating at each position angle while introducing Gaussian noise from the inverse variance image, we calculate an angle of maximal separation of these spatial centroids with an associated error.

Formally, the true position of maximal separation is given by PA$_{\mathrm{[OIII]}}$:
$$x_1 \cos(\mathrm{PA}_{\mathrm{[OIII]}}-\theta_2) = x_2 \cos(\mathrm{PA}_{\mathrm{[OIII]}} - \theta_1)$$

where $x_1$ and $x_2$ are the spatial separations at the observed position angles (Figure~\ref{a}), $\theta_1$ and $\theta_2$, respectively.

 \subsection{The Luminosity of the NLR}
 \label{size-luminosity}

To investigate the ionization structure of the NLR, we measure both the size and the [OIII]$\lambda 5007$ luminosity of the region. To determine the radius (in parsecs) of the NLR, we use half the Akaike width, defined in Section~\ref{kin setup}. We then convert to a physical distance using the Python astropy.cosmology utility. 

To determine the luminosity of the NLR, we use the SDSS DR7 OSSY value-added catalogue (\citealt{Oh2011}). We use a dereddened luminosity, L$_{\mathrm{[OIII]}}^c$, which is calculated from the observed [OIII] luminosity, L$_{\mathrm{[OIII]}}$, based upon a two component reddening correction (\citealt{Oh2011}). This includes a galaxy-wide dust correction as well as a nebular correction using the Hydrogen Balmer decrement:

$$\mathrm{L}_{\mathrm{[OIII]}}^c = \mathrm{L}_{\mathrm{[OIII]}} \Big(\frac{(\mathrm{H}\alpha/ \mathrm{H}\beta)_{\mathrm{obs}}}{3.0}\Big)^{2.94}$$
where $\mathrm{H}\alpha/\mathrm{H}\beta$ is the line ratio for the Balmer lines (\citealt{Osterbrock2006}).

The bolometric luminosity is calculated from L$_{\mathrm{[OIII]}}^c$:
$$\mathrm{L}_{\mathrm{bol}} = C \mathrm{L}_{\mathrm{[OIII]}}^c$$
where the correction, $C$, depends upon the [OIII] luminosity. $C$ is 87, 142, or 454 for L$_{\mathrm{[OIII]}}^c$ in the respective bins L$_{\mathrm{[OIII]}}^c$ (erg s$^{-1}$)$<10^{40}$, $10^{40}<\mathrm{L}_{\mathrm{[OIII]}}^c$ (erg s$^{-1}$)$<10^{42}$, and $10^{42}<\mathrm{L}_{\mathrm{[OIII]}}^c$ (erg s$^{-1}$)$<10^{44}$
(\citealt{Heckman2004}, \citealt{Lamastra2009}). We report L$_{\mathrm{[OIII]}}$, L$_{\mathrm{[OIII]}}^c$, the bolometric luminosity, and the radius of the NLR in parsecs in Table~\ref{R-L} for all galaxies.

\section{Analysis}
\label{analysis overall}
\subsection{Kinematic Classification}
\label{kin class}

We develop a novel technique for quantitative classification of the double-peaked [OIII]$\lambda 5007$ profiles using the spatially-resolved kinematics of the NLR. Our primary goal is to determine the nature of the double peaks and assess the relative importance of their various spectral features in the kinematic classification scheme. This technique depends on the longslit data alone. In future papers, the radio data will be independently analyzed and the two data sets will be synthesized.

We design the classification method to isolate rotation-dominated spectra from outflow-dominated spectra. \citet{MS2015} use a subsample of 18 galaxies to demonstrate that the majority (75\%) of double-peaked NLR galaxies are caused by `gas kinematics' (includes 70\% outflows and 5\% rotating NLR kinematics), 15\% are caused by dual AGNs or outflows produced by dual AGNs, and 10\% are ambiguous. \citet{Shen2011} and \citet{Fu2011} used resolved spectroscopy to show that the majority of double-peaked NLR galaxies are produced by `gas kinematics' from a single AGN, including extended emission-line nebulae, jet-cloud interactions, or peculiar narrow-line.
\begin{figure*}
	
\begin{center}
$\vcenter{\hbox{\includegraphics[clip, trim=0.5cm 2cm 1.5cm 3cm, width=0.35\textwidth]{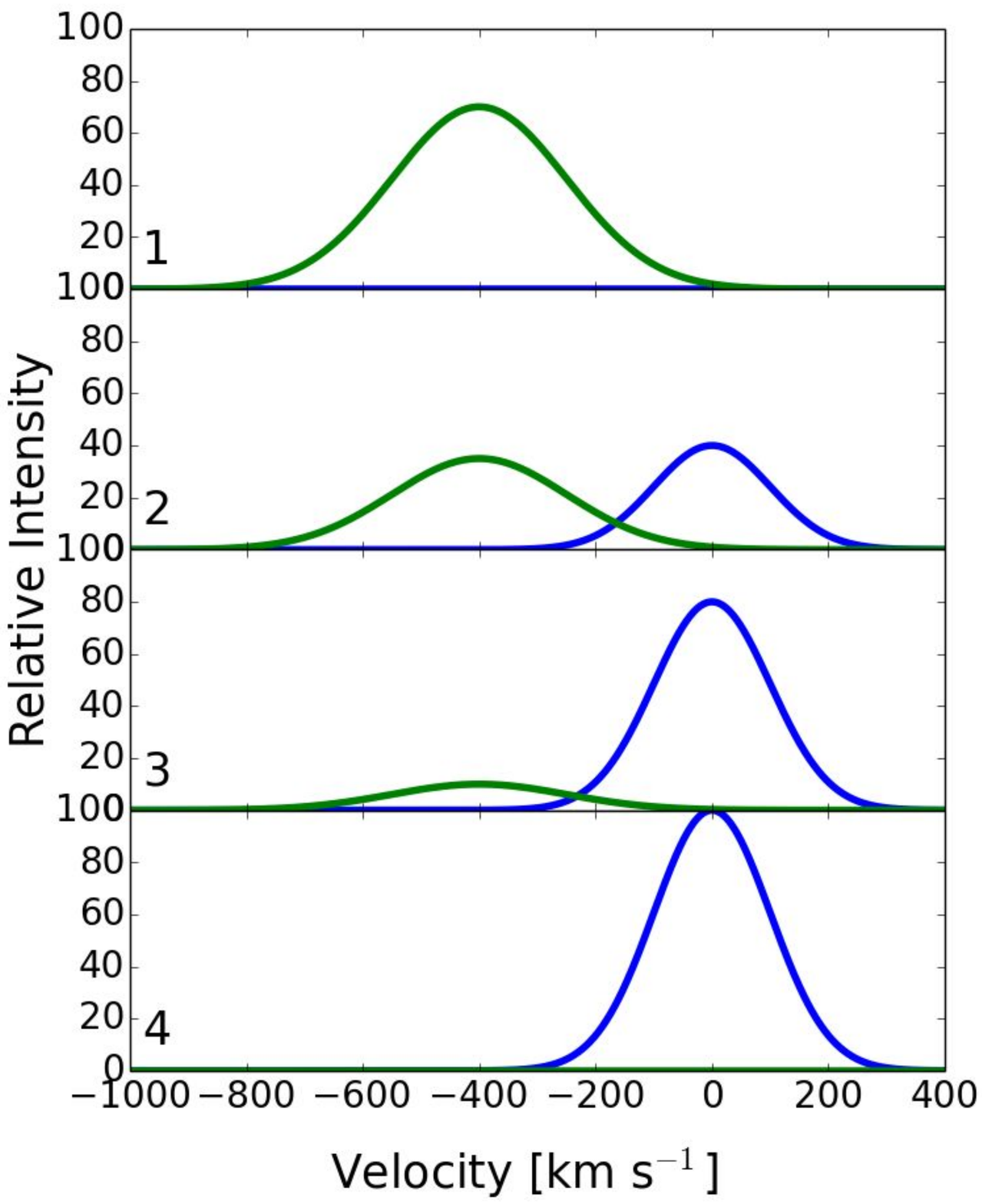}}}$
$\vcenter{\hbox{\includegraphics[clip, trim=2cm 9cm 2cm 9cm, width=0.50\textwidth]{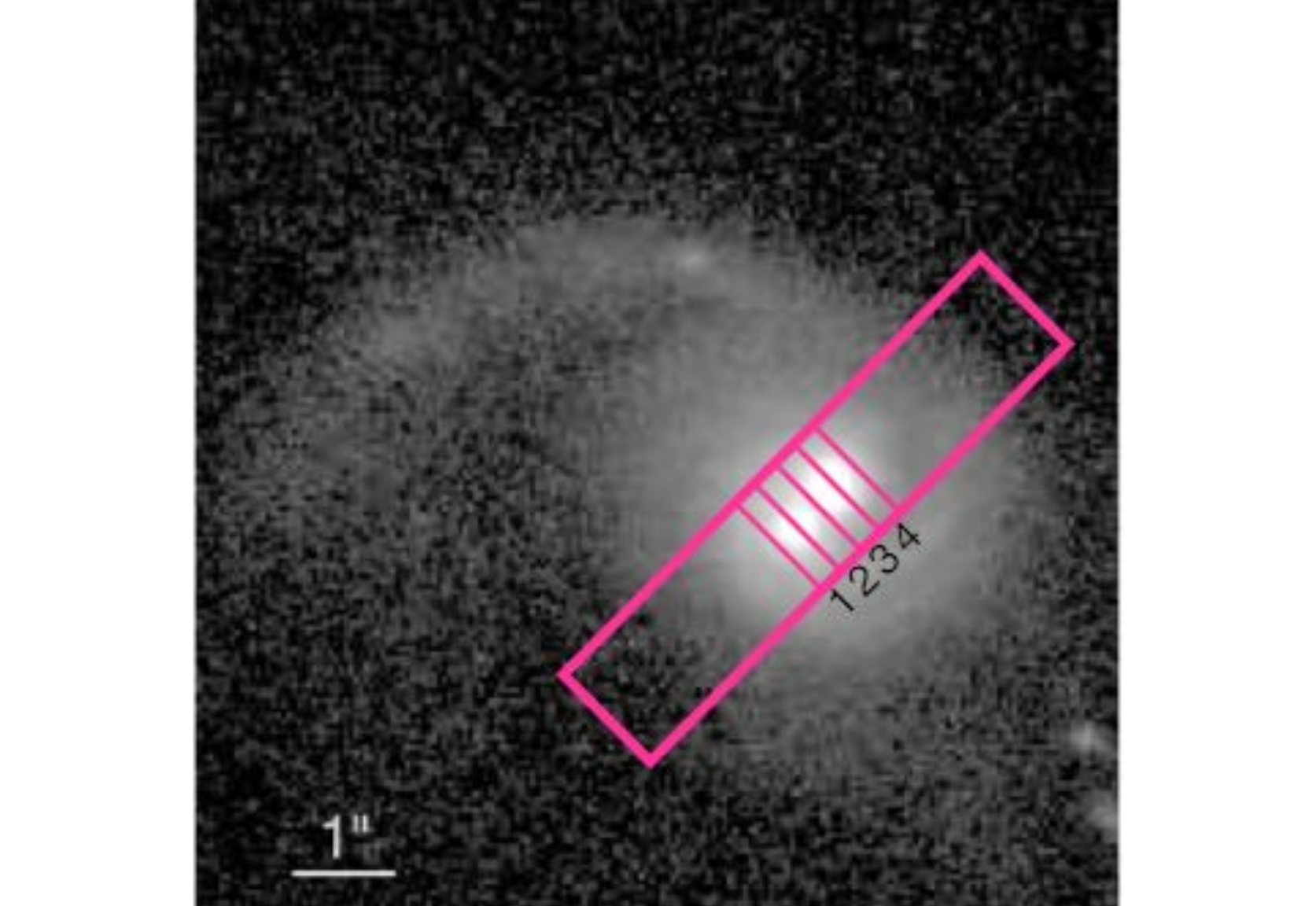}}}$
\caption{The spectrum of a toy model of a dual AGN that is characterized by the rotation-dominated NLRs of the dual AGNs with no outflow components. The x axis of the spatially-resolved spectrum (left) is in velocity space (km s$^{-1}$). On the right is the theoretical galaxy that hosts a dual AGN. The magenta box is the slit position (aligned SW to NE). We would expect to observe the longslit profiles as two distinct rotation-dominated NLRs with only one shifted peak at each spatial extremum and a double-peaked profile at the center. Note that based on the relative luminosity of the dual AGNs, the peaks may not be equal in flux as in this example. (Image taken from \citealt{Comerford2009b}). \label{dual AGN}}
\end{center}

\end{figure*}

\begin{figure*}
	
\begin{center}
$\vcenter{\hbox{\includegraphics[clip, trim=0.5cm 2cm 1.5cm 3cm, width=0.35\textwidth]{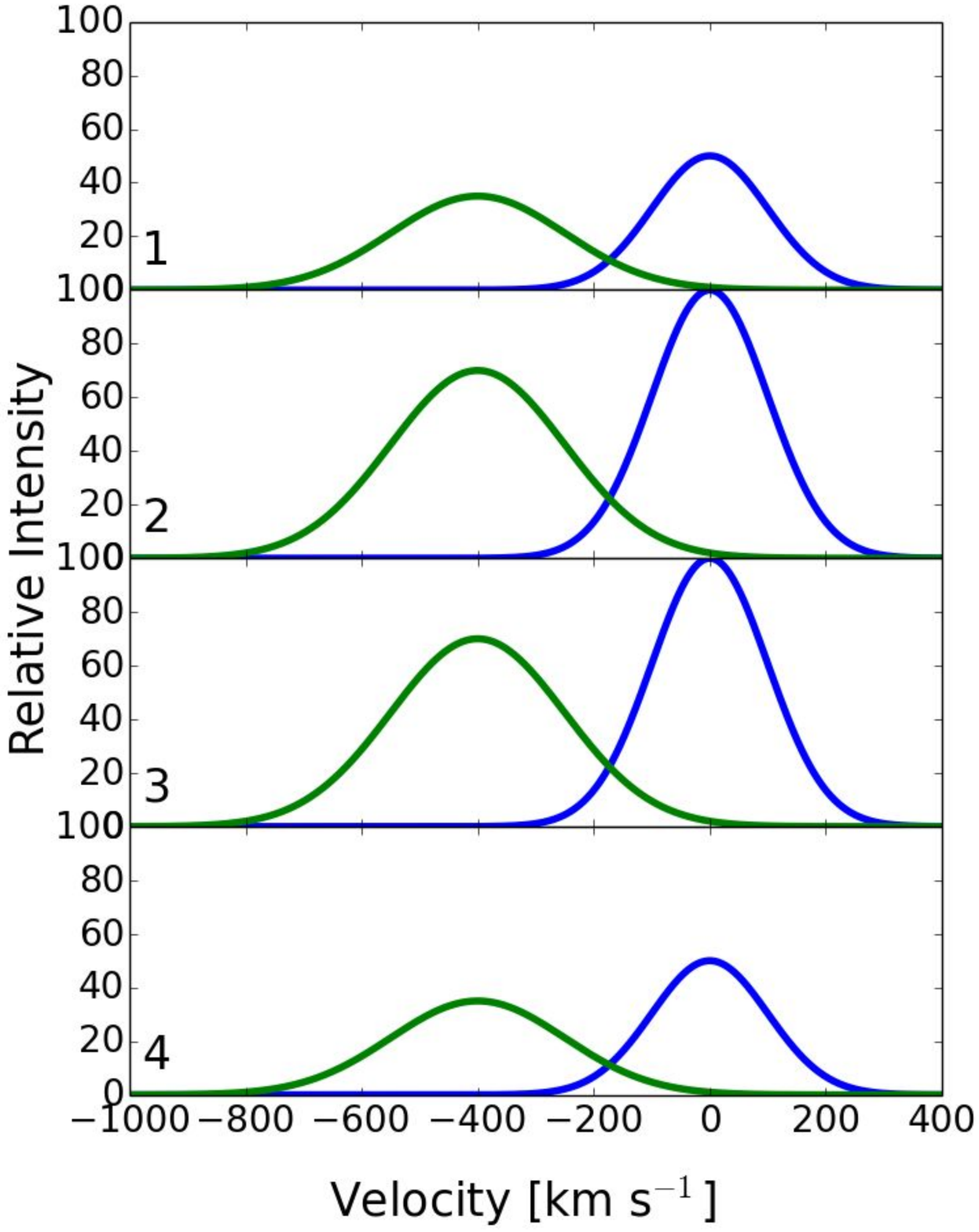}}}$
$\vcenter{\hbox{\includegraphics[clip, trim=2cm 9cm 2cm 9cm, width=0.50\textwidth]{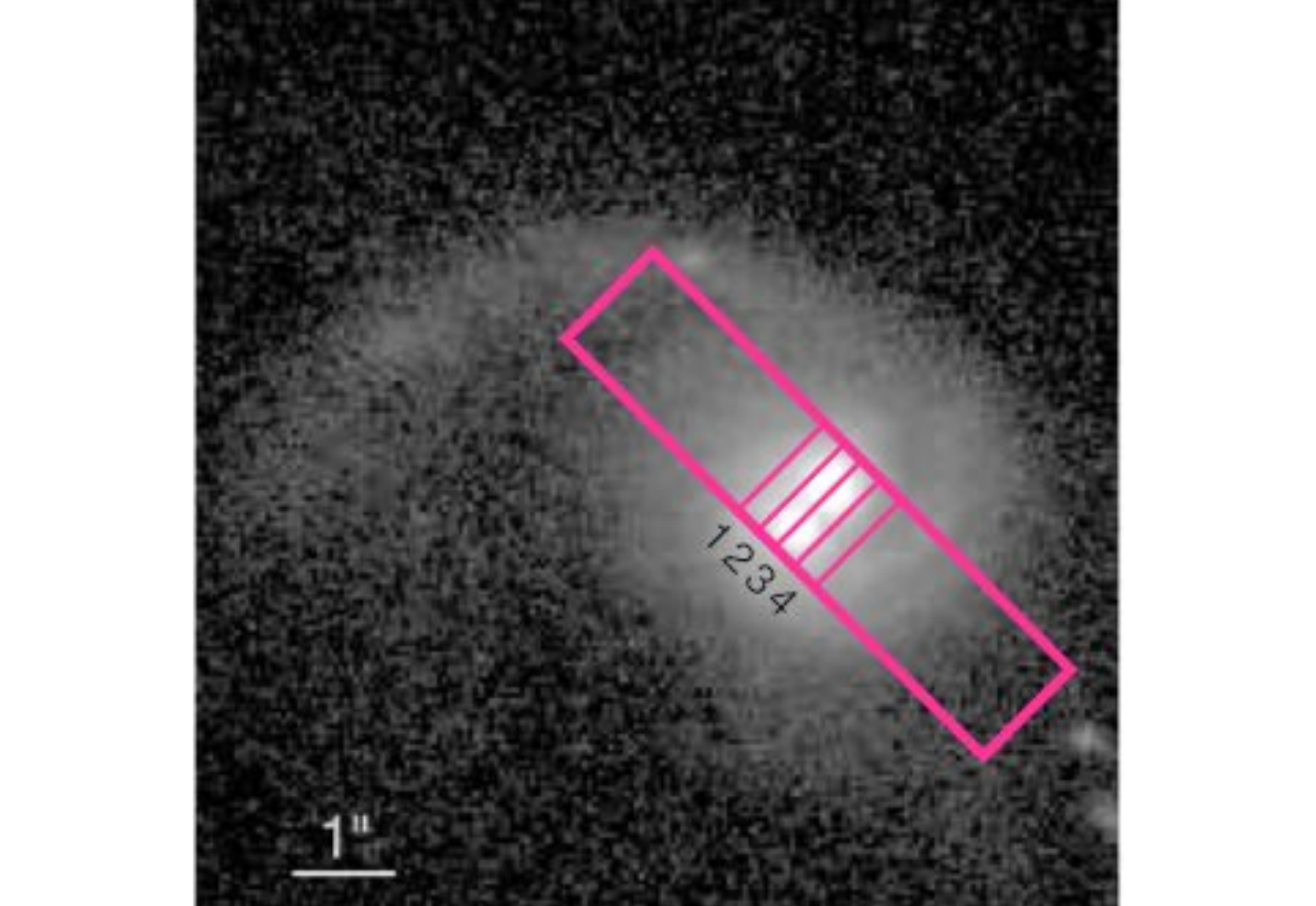}}}$
\caption{Same as Figure~\ref{dual AGN} but for the orthogonal slit position. At this PA we would observe the NLRs as overlapping because they are spatially coincident at all slit positions. Note that based on the relative luminosity of the dual AGNs, the peaks may not be equal in flux. The NLRs would be located at velocities centered at the relative velocities of the AGNs. (Image taken from \citealt{Comerford2009b}). \label{dual AGN PA 2}}
\end{center}

\end{figure*}

Since \citet{MS2015}, \citet{Shen2011}, and \citet{Fu2011} identify a statistical majority of outflow-dominated spectrum with the double-peaked selection technique, we create a classification technique that classifies a spectrum as outflow-dominated or rotation-dominated and then focuses on the kinematic nature of each broad classification. For instance, we further classify outflows as `Outflow Composite' if they have more than two Gaussian components or `Outflow' if they are best fit by two components. The `Outflow Composite' classification identifies outflows with complicated emission knots observed moving at distinct velocities. We further classify rotation-dominated spectra as containing a disturbance or an obscuration. 

Note that dual AGNs may exist in our classification scheme but fall into multiple different categories (Section~\ref{rot dual AGN})\footnote{The three dual AGNs from \citet{MS2015} are classified as Outflow Composite and Outflow, and are at higher redshifts ($z > 0.1$). They are classified as dual AGNs in \citet{MS2015} because they have two radio cores but these dual AGNs also have a powerful outflow component and their double-peaked profiles are well described as outflow-dominated.}. In this work, we explore the kinematic classifications for the complete sample of $z < 0.1$ double-peaked AGNs. Within this sample, there are no confirmed dual AGNs from the combination of radio and longslit data yet.

\begin{figure*}
\begin{center}
\includegraphics[scale=0.21]{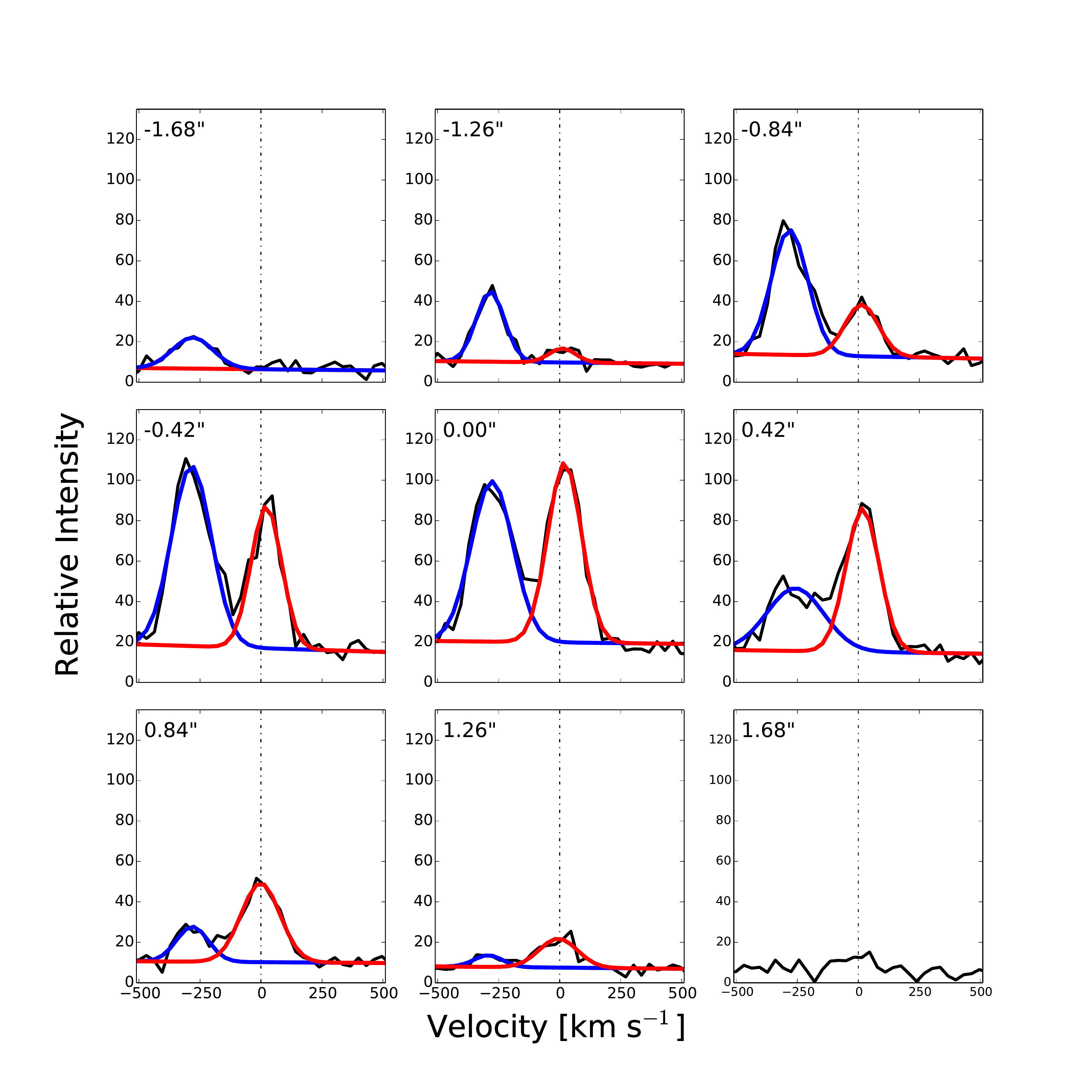}
\includegraphics[scale=0.21]{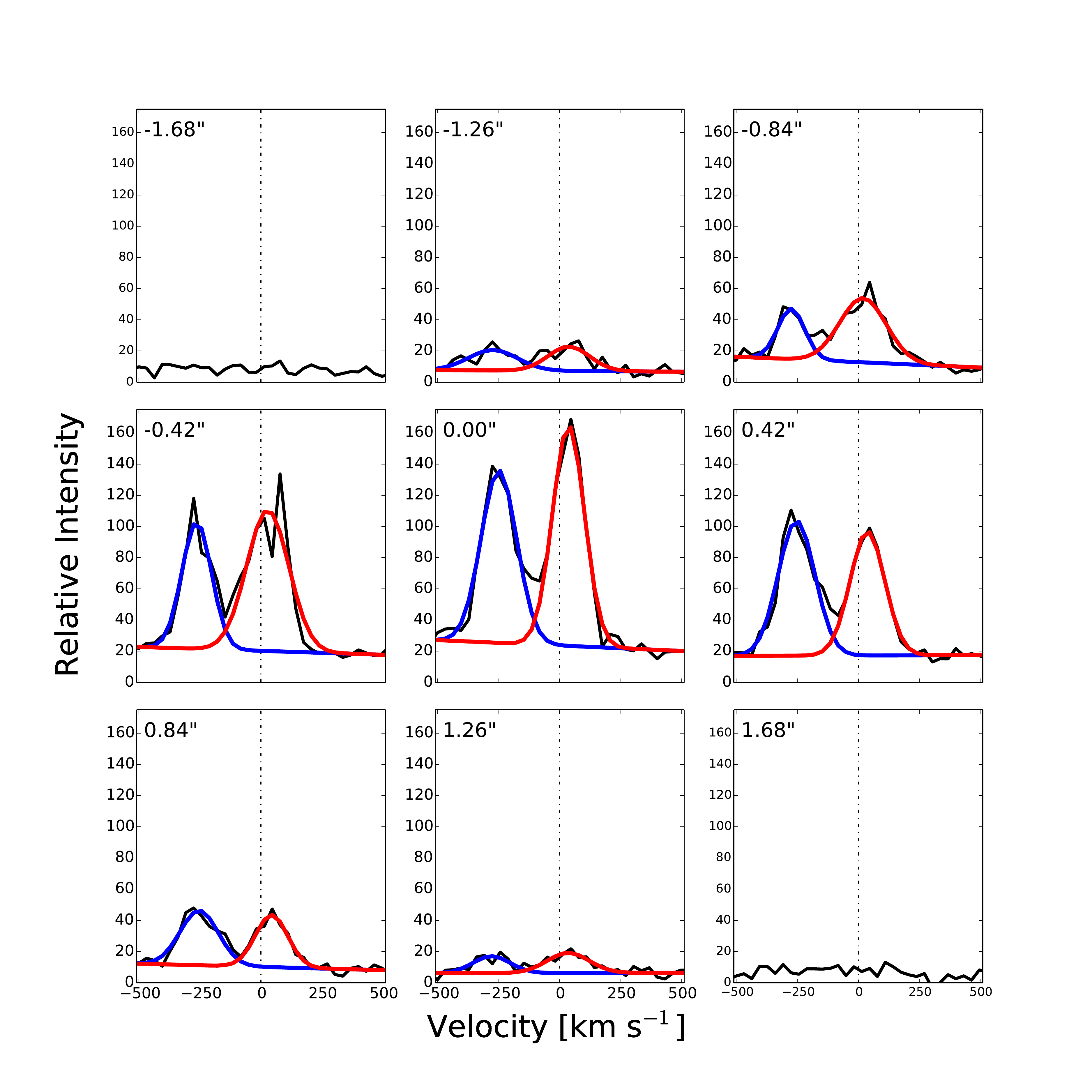}
\caption{Spatial profiles of emission for J1018+5127 at both PAs. We plot the spatial profiles for PA 22$^{\circ}$ (left) and 112$^{\circ}$ (right). On the left, the narrow components alternate in flux, matching the predictions for an outflow-free dual AGN at maximum spatial separation (Figure~\ref{dual AGN}). On the right, the narrow components do not alternate in flux, which is consistent with the prediction for a rotation-dominated dual AGN in Figure~\ref{dual AGN PA 2}. \label{J1018 both first}   }
\end{center}

\end{figure*}

In this work we will instead focus on the possible kinematic classifications of dual AGNs that will be confirmed in future work in which we synthesize the VLA radio data and the longslit kinematic data (M{\"u}ller-S{\'a}nchez et al. in prep.). In Figures~\ref{dual AGN} and~\ref{dual AGN PA 2} we explore the kinematic properties of a toy model of a dual AGN with no outflow components that is instead dominated by rotating NLR components. Each dual AGN has a rotating disk and both are orbiting in the potential of the host galaxy. We determine that one of the galaxies in the sample matches this observational prediction of a spectral profile of a rotation-dominated dual AGN. This galaxy, J1018+5127, is classified as `Rotation Dominated + Disturbance'. Figure~\ref{J1018 both first} shows the spatially-resolved profiles at both PAs for this candidate dual AGN. However, since this candidate lacks confirmation of dual radio cores, we do not create a separate kinematic classification for `dual AGNs' and instead allow candidate dual AGNs to fall under different kinematic classifications that describe the nature of the dual AGNs. Additionally, other work has predicted that dual AGNs such as the toy model presented here that are characterized purely by rotation-dominated NLRs are not as common as dual AGNs with outflow components (\citealt{Blecha2013}).

\begin{deluxetable*}{lllllllll}[h]

\tabletypesize{\scriptsize}
\tablewidth{0pt}
\tablecolumns{9}
\tablecaption{Kinematic Classification Part 1}
\tablehead{
\colhead{SDSS ID} & 
\colhead{PA$_{\mathrm{obs}}$} & 
\colhead{Number of Rows} & 
\colhead{Single Gauss} & 
\colhead{Single Gauss} & 
\colhead{Double Gauss} & 
\colhead{Double Gauss} & 
\colhead{PA$_\mathrm{{gal}}$} & 
\colhead{PA$_\mathrm{[OIII]}$} \\
& & with $>2$ Gaussian & V$_r$ [km $\mathrm{s}^{-1}$] & $\sigma$ [km $\mathrm{s}^{-1}$] & 
$\sigma_1$ [km $\mathrm{s}^{-1}$] & $\sigma_2$ [km $\mathrm{s}^{-1}$] & & \\
& & Components & & & & & & 
}
\startdata

J0002+0045 & 68 &5/9        & $ 173.8 \pm 3.3 $       & $1005.0 \pm 8.7 $ &$ 655.8 \pm 198.0 $ &$ 920.2 \pm 59.8          $         & 65              & 64                         \\
                       
                                  &    158 &   4/7       & $ 84.9 \pm 0.8 $           & $1018.5\pm 9.6$   & $408.5 \pm 3.3    $ & $371.7 \pm 1.2   $  \\

J0009$-$0036  &23 & 5/11          & $ 144.4 \pm 5.1   $       & $650.0 \pm 4.3 $  & $736.2 \pm 6.8$    & $624.1 \pm 206.7 $       & 135             & 51                          \\
                               &  67  &  0/11        &$ 235.6 \pm 1.9 $         & $619.4 \pm 3.0 $  & $705.7 \pm 8.7    $ & $610.7 \pm 191.3$   \\

J0135+1435 &20 & 2/7 & $ 218.0 \pm 0.9 $  & $536.5 \pm 173.4 $& $469.0 \pm 169.7$ & $298.9 \pm 123.8$ &20 &46 \\
&81&  3/5& $ 78.2 \pm 0.5 $  & $500.5 \pm 3.7$ & $515.1 \pm 4.6$ & $285.9 \pm 174.7$ \\

J0156$-$0007 &29&1/5 & $140.0 \pm 1.0$  & $672.5 \pm 282.6$ & $698.3 \pm 358.9$ &$ 527.3 \pm 227.0$ &119 &113\\
&119& 3/3& $241.5 \pm 3.0$  & $618.4 \pm 6.2$ & $419.7 \pm 20.5 $& $616.4 \pm 217.3$ \\

\enddata

\label{tabkinnum}
\tablecomments{The values for the parameters for quantitative classification for both observed PAs. Column 1: Galaxy name. Column 2: Observed position angle. Column 3: The number of rows that are best fit by $>2$ Gaussians within the resolved center of emission (FWHM$_{\mathrm{cont}} \pm$1 row). For instance, in row one, 5/9 indicates that out of the nine rows of the resolved center of emission, 5 rows are better fit by 3 Gaussians. Column 4: V$_r$ refers to the line of sight velocity from the single Gaussian centroid. Column 5, 6, and 7: The values for the dispersion and error of one component, and each individual component for the two Gaussian fit. Column 8 and 9: We create a diagnostic for alignment by listing the position angles of the photometric major axis and the [OIII] emission. PA$_{\mathrm{gal}}$ is from SDSS $r$-band photometry.
(This table is available in its entirety in the online journal.)}

\end{deluxetable*}

\begin{figure*}[h]
\begin{center}
\includegraphics[clip, trim=2cm 0.0cm 0.0cm 0.0cm, width=1.1\textwidth]{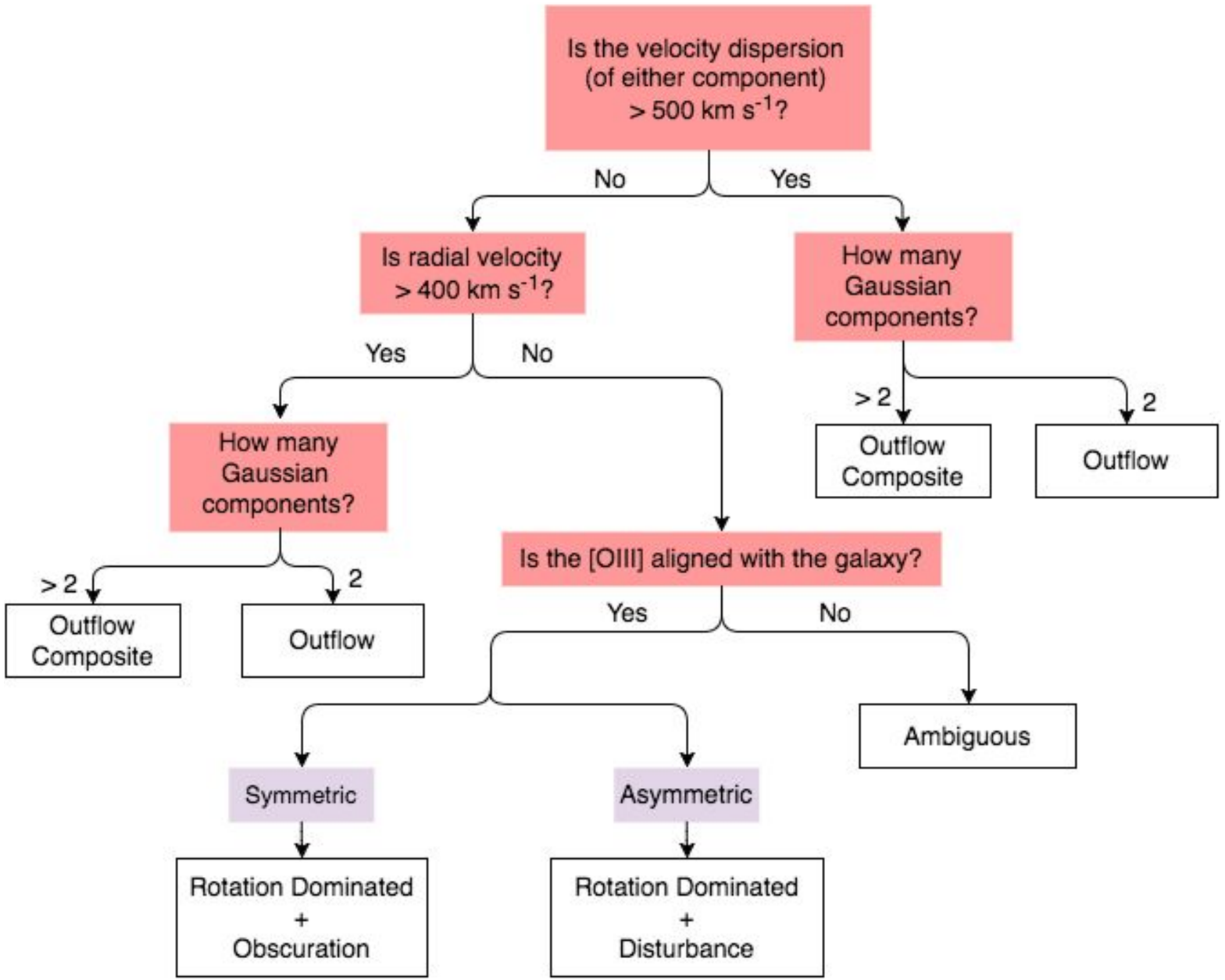}
\caption{Kinematic classification mapping scheme. We demonstrate the sequence of classifications that occur for each galaxy. The classification terminates in either an outflow-dominated classification (Outflow Composite or Outflow), a rotation-dominated classification (Rotation Dominated + Obscuration or Rotation Dominated + Disturbance), or `Ambiguous' (some combination of inflow, outflow, and counter-rotating disk). \label{classify}}
\end{center}

\end{figure*}

We find that the most powerful quantitative properties for identifying the kinematic nature of a spectral profile are the velocity dispersion of each of the individual Gaussians in a two Gaussian fit ($\sigma_1$ and $\sigma_2$), the radial velocity of a one Gaussian fit (V$_r$), the number of kinematic components, and the alignment of the [OIII]$\lambda 5007$ emission with the major axis of the galaxy. Figure~\ref{classify} demonstrates the classification scheme based upon these parameters, Table~\ref{tabkinnum}  shows the values of these properties for both observed position angles of each galaxy, and Table~\ref{tabkin} produces the final classification for the entire sample of 71 galaxies.

\begin{deluxetable*}{lllllll}

\tabletypesize{\scriptsize}
\tablewidth{0pt}
\tablecolumns{7}
\tablecaption{Kinematic Classification Part 2}
\tablehead{
\colhead{SDSS ID} & 
\colhead{$>2$ Gaussian} & 
\colhead{V$_r$ [km $\mathrm{s}^{-1}$]} & 
\colhead{$\sigma$ [km $\mathrm{s}^{-1}$]} & 
\colhead{$\sigma_1$ and $\sigma_2$ } & 
\colhead{Aligned?} & 
\colhead{Classification} \\
& Components?& & & [km $\mathrm{s}^{-1}$] & PA$_{\mathrm{[OIII]}}$=PA$_{\mathrm{gal}}$ & 
}

\startdata
J0002+0045&   $>2$&  $<400$&  $>500$  &  $>500$    &yes & Outflow Composite\\
J0009$-$0036&2 &$<400$&$>500$  &$>500$    &no&Outflow\\
J0135+1435&$>2$ & $<400$& $\geq 500$ & $> 500$ & no  & Outflow Composite  \\
J0156$-$0007& $>2$ & $<400$& $>500$ & $\geq 500^a$& yes  & Outflow Composite \\

\enddata

\tablecomments{Table of the official kinematic classification for each galaxy based upon the measurements in Table~\ref{tabkinnum}. Column 1: Galaxy name. Column 2: The number of components is $>2$ if at either position angle, more than half the rows within the spatial center of the galaxy have $>2$ components as the best fit. Column 3: V$_r$ is maximum line of sight radial velocity, measured from the single Gaussian fit for both position angles. Column 4: $\sigma$ is given by the largest dispersion of the single Gaussian fit from both observed position angles. We include this measurement in the table because although it is not used in the classification scheme, it is discussed in Section~\ref{results kin class}. Column 5: Likewise, $\sigma_1$ or $\sigma_2$ is the largest dispersion of either of the two Gaussian fit fits from both position angles. Column 6: Alignment is determined from PA$_{\mathrm{[OIII]}}$ and PA$_{\mathrm{gal}}$. If these two measurements are within 20$^{\circ}$ of one another, the galaxy is aligned. Column 7: We provide a classification based upon these five columns and four properties. 
(This table is available in its entirety in the online journal.)}
\tablenotetext{a}{The 1$\sigma$ error on the measured value straddles the classification cutoff. However, the measured value  greater than 500 km s$^{-1}$ so we classify this galaxy as Outflow Composite. For all other galaxies in our sample in which the measured value straddles a given classification cutoff within error are classified according to the measured value.} 
\label{tabkin} 
\end{deluxetable*}

Using these four properties of the [OIII]$\lambda 5007$ profiles, we first separate rotation-dominated profiles from outflow-dominated profiles. This elimination-based technique is useful because the observational properties of rotating structure have more stringent constraints. For instance, rotating structure behaves according to Keplerian physics, which place limits on the line of sight velocity (V$_r <$ 400 km $\mathrm{s}^{-1}$) and velocity dispersion ($\sigma_1$ and $\sigma_2 < 500$ km $\mathrm{s}^{-1}$, \citealt{Osterbrock2006}). These observationally-defined velocity limits are used to divide outflow-dominated from rotation-dominated kinematics.

\begin{figure*}
\begin{center}
\includegraphics[clip, trim=0.0cm 0.0cm 0.0cm 8cm, width=0.80\textwidth]{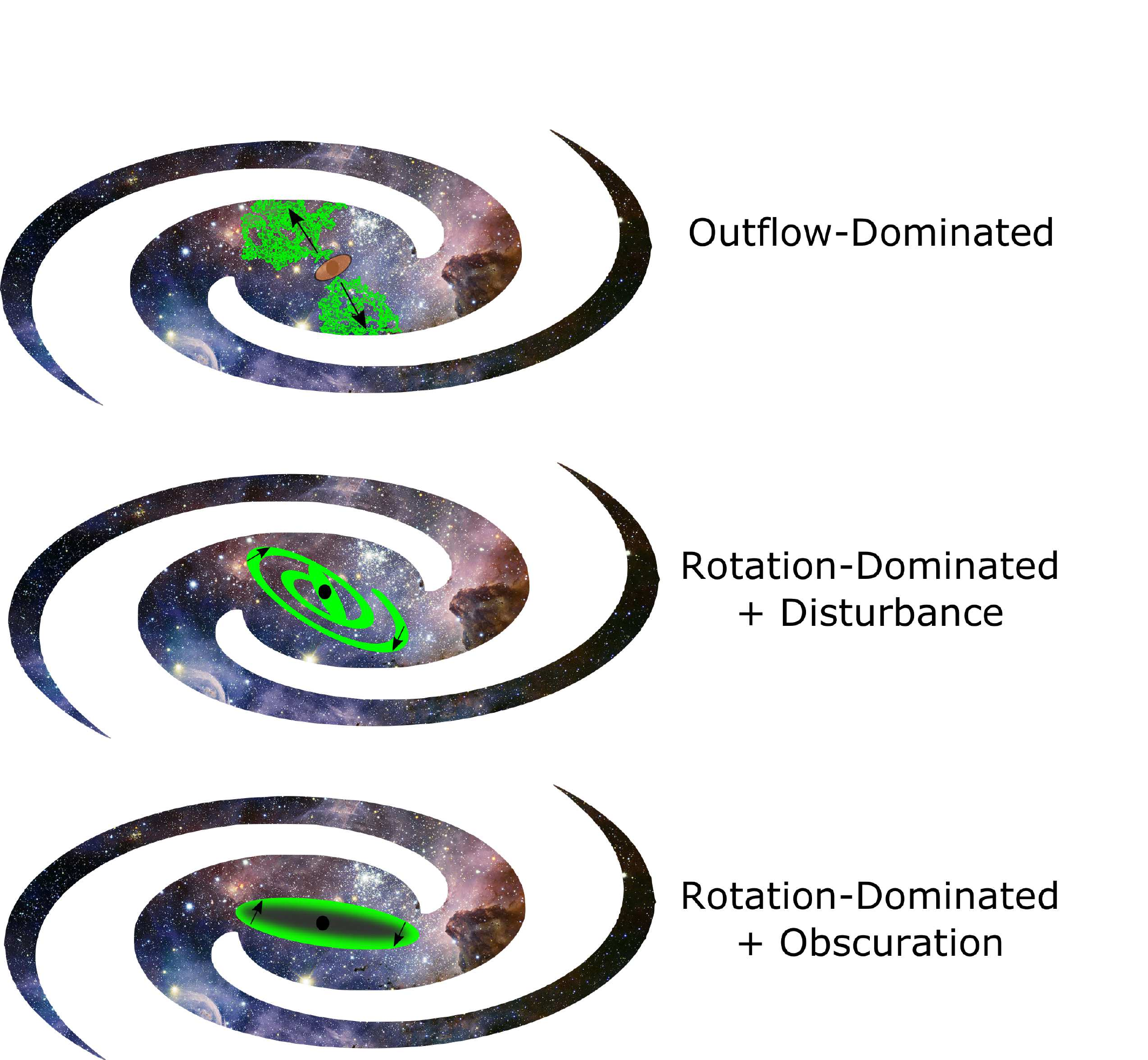}

\caption{Cartoon diagrams of three of the main kinematic classifications. Here we show the origin of the [OIII]$\lambda 5007$ kinematics in green for an outflow-dominated classification and the two rotation-dominated classifications. We demonstrate the kinematic components that are responsible for the velocity shifts in the [OIII]$\lambda 5007$ profiles using black arrows. Top: An outflow-dominated galaxy, `Outflow' or `Outflow Composite', is characterized by outflowing [OIII]$\lambda 5007$ emission. Middle: A Rotation Dominated + Disturbance galaxy may have rotating spiral structure or a bar structure. Bottom: A Rotation Dominated + Obscuration galaxy is characterized by a rotating disk that has a co-rotating obscuring structure such as a dust lane that produces a double-peaked profile at each spatial position. \label{oiiigals}}
\end{center}

\end{figure*}

We identify outflow-dominated profiles as galaxies with velocity dispersions and line of sight velocities that are in excess of the limits given above for rotation-dominated profiles. The presence of a broad component ($\sigma_1$ or $ \sigma_2 > 500$ km s$^{-1}$) demonstrates the presence of an outflow (\citealt{MS2011}). Likewise, line of sight velocities that exceed 400 km s$^{-1}$ (V$_r > 400$ km s$^{-1}$) for a single Gaussian fit identify an outflow because discrete knots of emission have been observed at these velocities in outflows (e.g., \citealt{Das2006}, \citealt{Fischer2013}, \citealt{CF2015}). These outflow-dominated galaxies are then further classified into `Outflow' or `Outflow Composite' according to the number of kinematic components. For instance, a profile with $>2$ kinematic components is designated `Outflow Composite' (determined using Akaike statistics described in Section~\ref{methods}). Outflows can show $>2$ components because they have distinct clouds of gas that move at a variety of discrete velocities dominated by a central engine. For a diagram of an outflow-dominated galaxy, see Figure~\ref{oiiigals}.

Galaxies that are not classified as outflow-dominated continue along the quantitative classification scheme; we next determine if a galaxy is rotation-dominated or ambiguous according to the alignment. Alignment of the ionized gas with the stellar disk is a property of rotation-dominated galaxies. We measure alignment by comparing PA$_{\mathrm{gal}}$, the position angle of the photometric major axis of the galaxy from the SDSS $r$-band, to PA$_{\mathrm{[OIII]}}$ (measured in Section~\ref{methods}). As discussed in \citet{Comerford2012}, a rotationally dominated double-peaked NLR should be aligned with the plane of the galaxy because the galaxy's potential dominates the kinematics; thus for these galaxies, PA$_{\mathrm{[OIII]}} \sim \mathrm{PA}_{\mathrm{gal}}$ within a $20^{\circ}$ error (as in \citealt{MS2015}). If the [OIII]$\lambda 5007$ emission is aligned with the plane of the galaxy, the emission is classified as rotation-dominated. 

If the emission is not aligned with the plane of the galaxy and the emission has not already been classified as an outflow according to the value of radial velocity or the individual velocity dispersions, the emission could be a counter-rotating disk, an outflow, an inflow, or some combination of these kinematic origins. We could further tie the gas kinematics to an inflow origin if the galaxy were undergoing a merger since mergers funnel gas to the center of the galaxy (e.g., \citealt{Hopkins2006}). However, these gas kinematics could be also be explained by an outflow or counter-rotating disk in a merger. If the galaxy is not undergoing a merger, we can rule out merger-driven inflows for the kinematic origin of the disturbed NLR kinematics. However, it would still be difficult to distinguish between an outflow-dominated kinematic origin to the NLR or a counter-rotating disk. We classify these galaxies as `Ambiguous' since we do not have the ability to fully determine the presence of a merger in these galaxies based upon SDSS imaging alone. Additionally, in either case, this type of galaxy is still ambiguous in its classification.

We use alignment as a classification tool once we have already classified the outflow-dominated galaxies; therefore, galaxies with outflows may also have emission that is aligned with the photometric major axis of the galaxy. This will be important in later work where we discuss the implications of the geometry of the outflow-dominated galaxies (Nevin et al. in prep.). Likewise, the number of Gaussian components is not used to further classify rotation-dominated galaxies. A rotation-dominated galaxy may have more than two Gaussian components, but we do not use this to further classify rotation-dominated galaxies into subcategories.

Within the category of rotation-dominated spectra, we further classify galaxies as `Rotation Dominated + Obscuration' or `Rotation Dominated + Disturbance' (such as a bar or spiral). If a galaxy has all of the kinematic properties of a rotation-dominated galaxy, the [OIII]$\lambda 5007$ emission is aligned with the kinematic major axis of the galaxy, and the galaxy has a symmetric profile (compared to both the full sample and the rotation-dominated galaxies), we classify it as Rotation Dominated + Obscuration. \citet{Smith2012} suggest that equal flux double-peaked symmetric profiles are rotating disks. We classify one galaxy (J0736+4759) from the sample of 71 galaxies as Rotation Dominated + Obscuration due to its high degree of symmetry. Radiative transfer effects from a central dust lane could account for a single peaked Gaussian with a decrease in flux at zero velocity at both observed position angles, which would produce a double-peaked profile. Figure~\ref{oiiigals} shows a diagram of this type of kinematic origin of a double-peaked profile.

Dynamic disturbances in the plane of the galaxy could also account for a rotation-dominated spectrum with a double-peaked profile. We classify asymmetric rotation-dominated profiles where the [OIII]$\lambda 5007$ emission is aligned with the kinematic major axis as `Rotation Dominated + Disturbance'. These galaxies could host nuclear bars, spiral arms, or a kinematically disturbed dual AGN that causes asymmetric double-peaked profiles (\citealt{Davies2009,Hicks2009,Schoenmakers1997,Blecha2013}). Bars or spirals accelerate the zero velocity gas through infall or chaotic motion (\citealt{MS2009}). This enhances the wings of the single peaked rotation-dominated profile and produces a double peaked profile that is asymmetric in flux. Again, Figure~\ref{oiiigals} shows a diagram of a disturbed rotation-dominated galaxy. Note that distinguishing between a disturbed and obscured rotation-dominated NLR is the only category of classification that requires the asymmetry parameter. We discuss the quantitative determination of relative asymmetry for the rotation-dominated galaxies in Section~\ref{rot analysis}.

\begin{figure*}
\begin{center}
\includegraphics[scale=0.21]{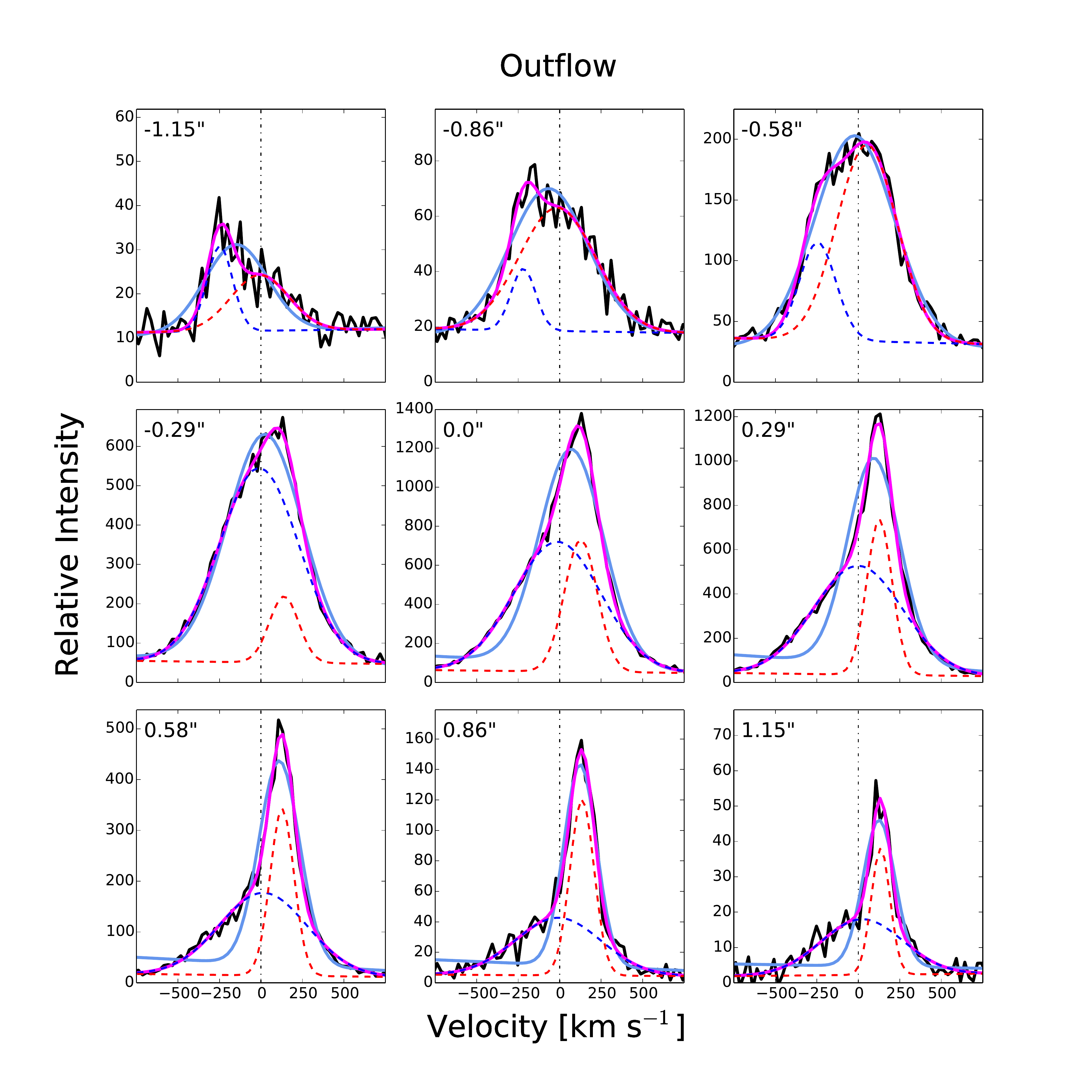}
\includegraphics[scale=0.21]{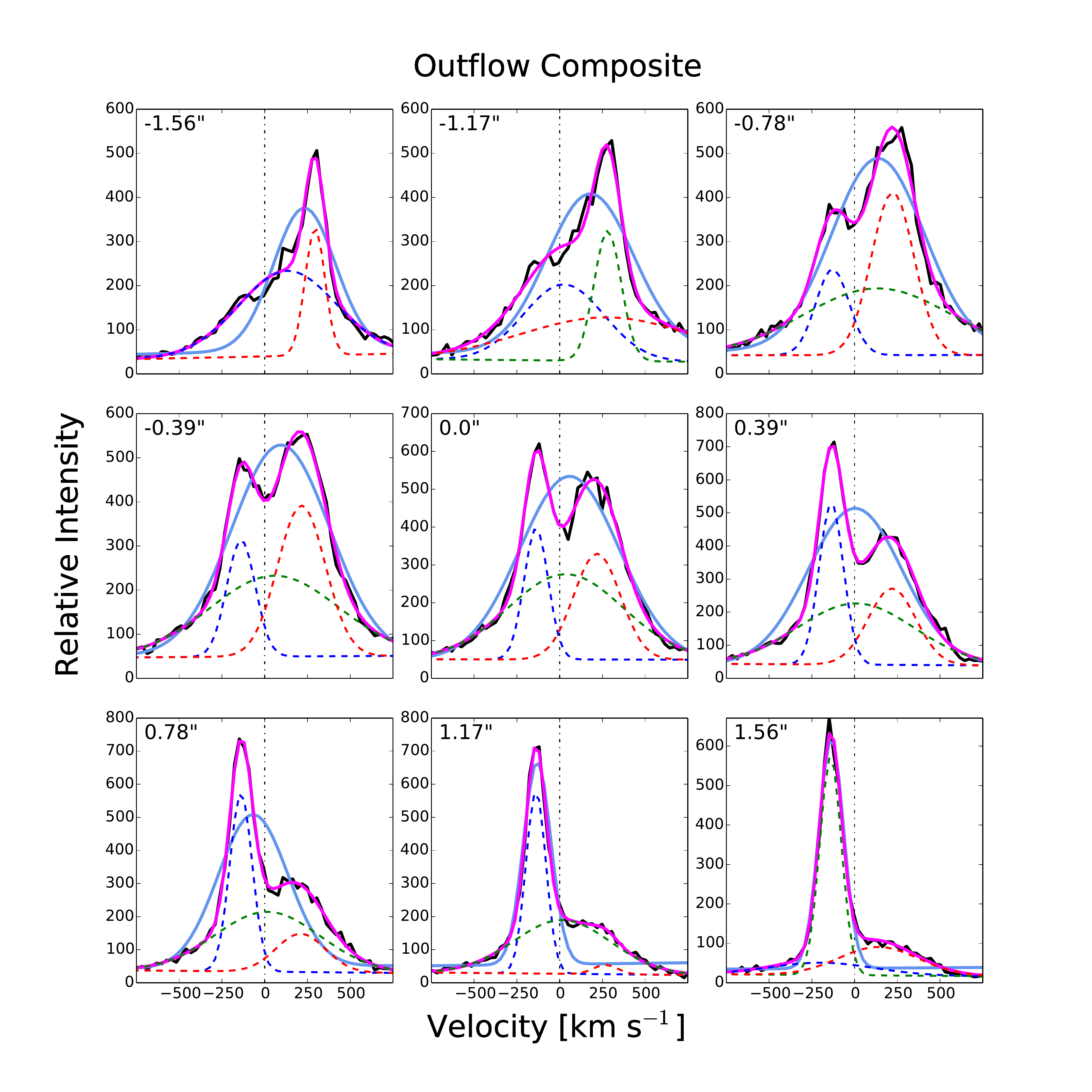}
\caption{Examples of galaxy profiles for galaxies classified as `Outflow' (left) and `Outflow Composite' (right). We plot the profiles for the nine central spatial rows, with the spatial position shown in arcseconds in the upper left corner of each panel. The data are in black and the integrated one Gaussian fit is a blue solid line. Here we plot a two Gaussian fit for the galaxy classified as `Outflow' (left), where the red and blue dashed lines are the individual Gaussian components and the solid magenta line is the integrated two Gaussian fit. For the `Outflow Composite' galaxy, if a given spatial row is better fit by a three Gaussian fit, we plot the integrated fit for the three Gaussian fit where the dashed red, green, and blue lines are the individual Gaussian component fits and the solid magenta line is the integrated three Gaussian fit. For rows where two Gaussians are a better fit, we plot the individual and integrated profiles from the two Gaussian fit. We plot the three Gaussian fits for the `Outflow Composite' galaxy to demonstrate that three components are sometimes necessary to fit the low flux wider wings of this type of profile. Here, we identify these galaxies as outflow-dominated due to the velocity dispersions ($\sigma_1$ or $\sigma_2 > 500$ km s$^{-1}$) of the individual components of the two Gaussian fit. We then further classify one galaxy as `Outflow Composite' because three Gaussian components are a better fit for more than half the spatial rows. Conversely, the `Outflow' galaxy is better fit with two Gaussian components. \label{proofconcept1}}
\end{center}

\end{figure*}

\begin{figure*}
\begin{center}
\includegraphics[scale=0.3]{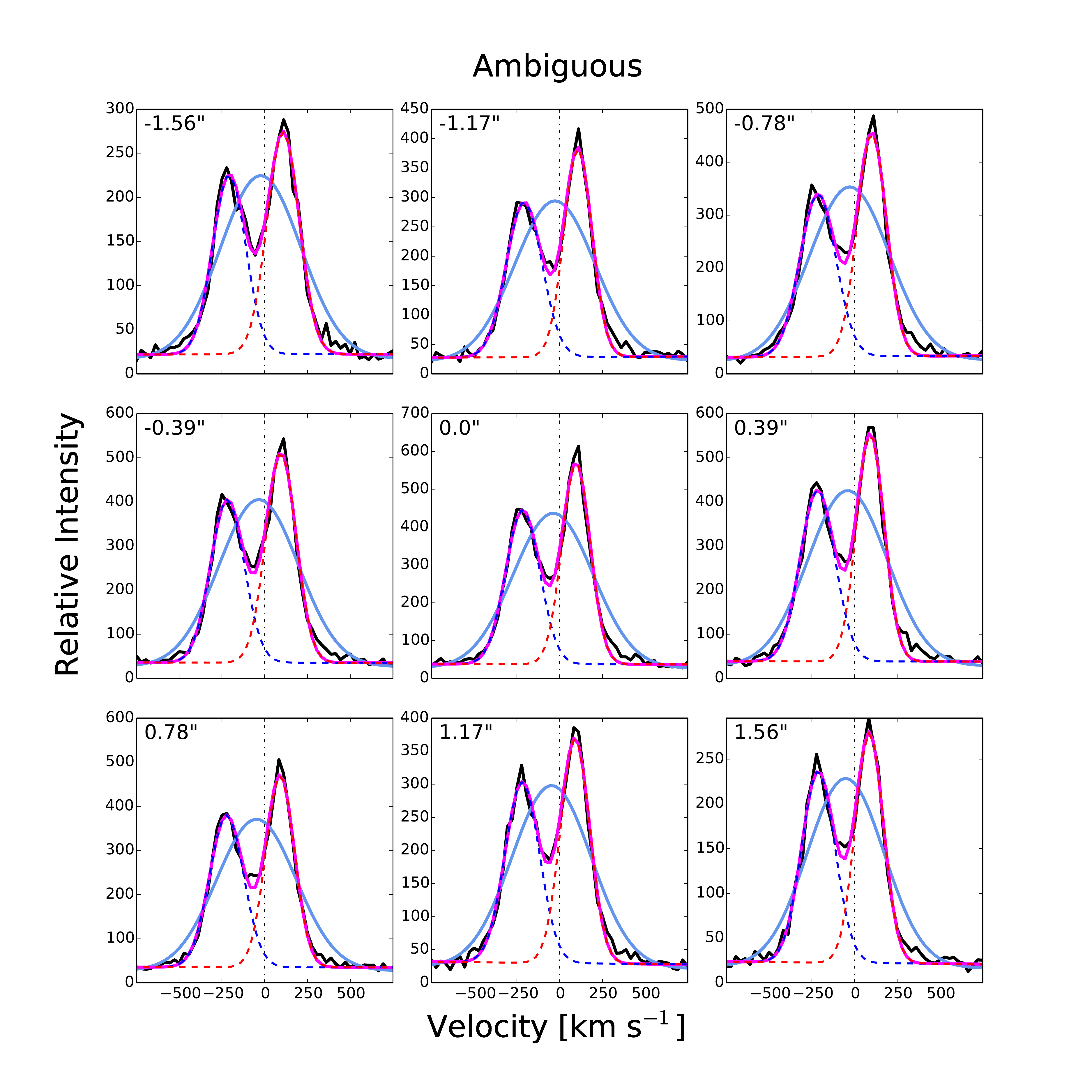}
\includegraphics[scale=0.3]{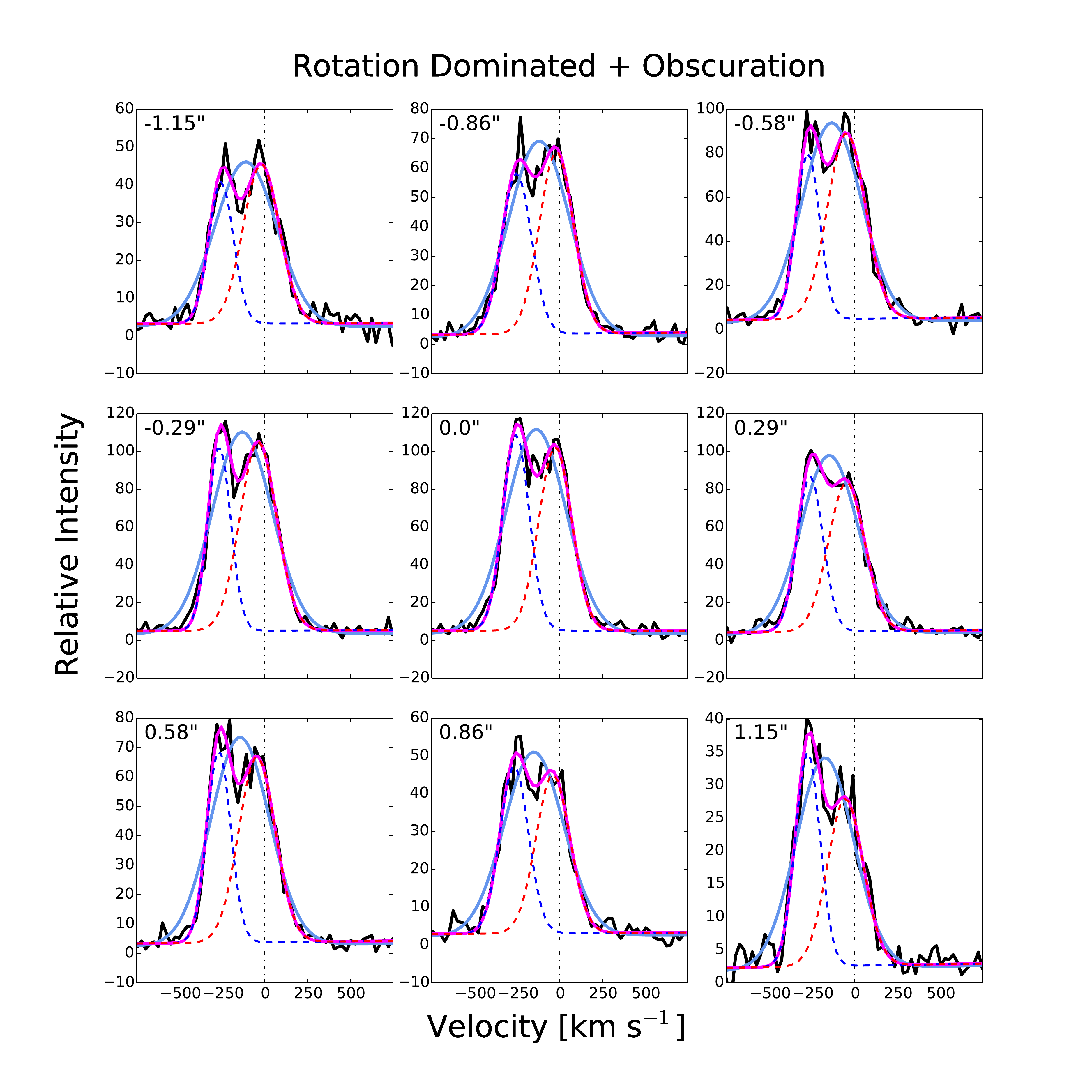}
\end{center}	
\end{figure*}

\begin{figure*}
\begin{center}
\includegraphics[scale=0.3]{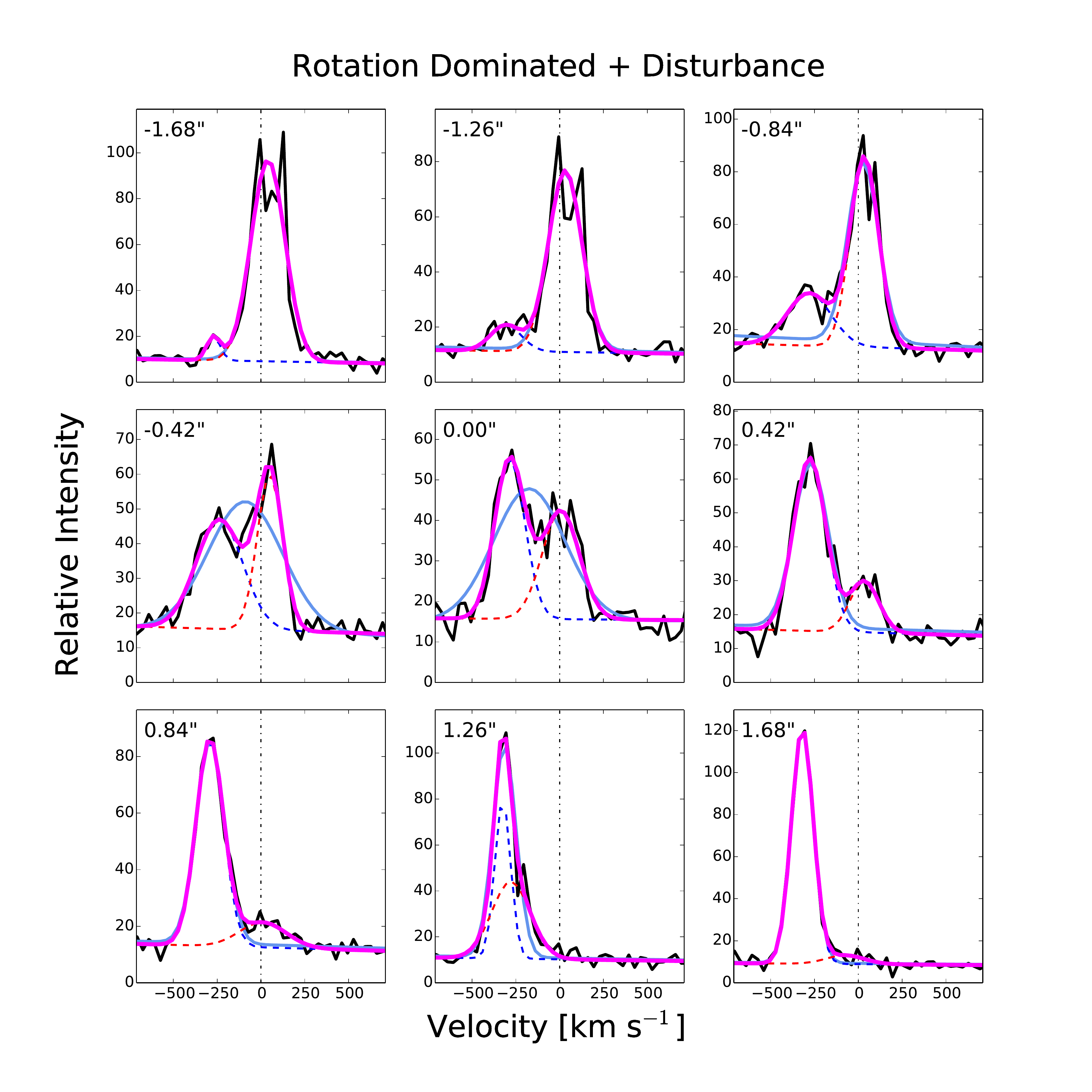}

\caption{Same as Figure~\ref{proofconcept1}, but for the classifications of `Ambiguous', `Rotation Dominated + Obscuration', and `Rotation Dominated + Disturbance'. We show the individual Gaussian components of the two Gaussian fits to demonstrate our classification technique's ability to identify the velocity dispersion and velocity offset of each of these components. All three galaxies are excluded from our outflow-dominated classification because the individual velocity dispersions are low ($\sigma_1$ and $\sigma_2 < 500$ km s$^{-1}$) and the line of sight velocity is low (V$_r < 400$ km s$^{-1}$). These three galaxies are then classified based upon their asymmetry or the alignment of the [OIII] emission (Section~\ref{rot analysis}). The galaxy classified as `Ambiguous' has profile characteristics that are consistent with a rotation-dominated galaxy; the only characteristic that sets it apart is the fact that the [OIII] emission is not aligned with the photometric major axis of the galaxy. See Figure~\ref{Alignment} for a visualization of the [OIII] alignment of the galaxies classified here as `Ambiguous' and `Rotation Dominated + Obscuration'. The galaxy classified as `Rotation Dominated + Obscuration' has [OIII] emission that is aligned with the photometric major axis of the galaxy and a symmetric profile. The galaxy classified as `Rotation Dominated + Disturbance' has aligned [OIII] emission and an asymmetric profile. \label{proofconcept}}
\end{center}

\end{figure*}

\begin{figure*}
\begin{center}
$\vcenter{\hbox{\includegraphics[clip, trim=2cm 5cm 2cm 5cm, width=0.4\textwidth]{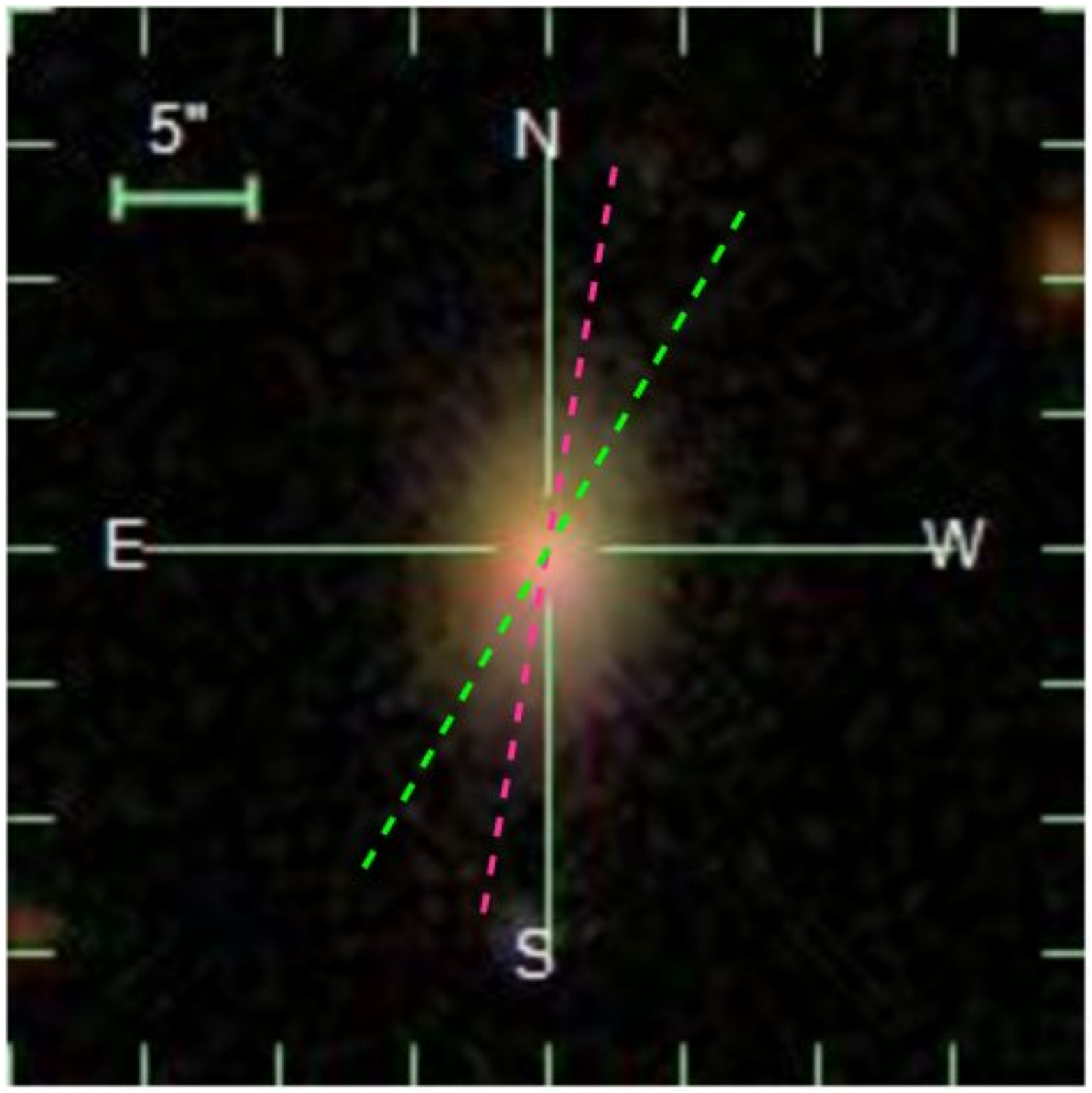}}}$
$\vcenter{\hbox{\includegraphics[clip, trim=2cm 5cm 2cm 5cm, width=0.5\textwidth]{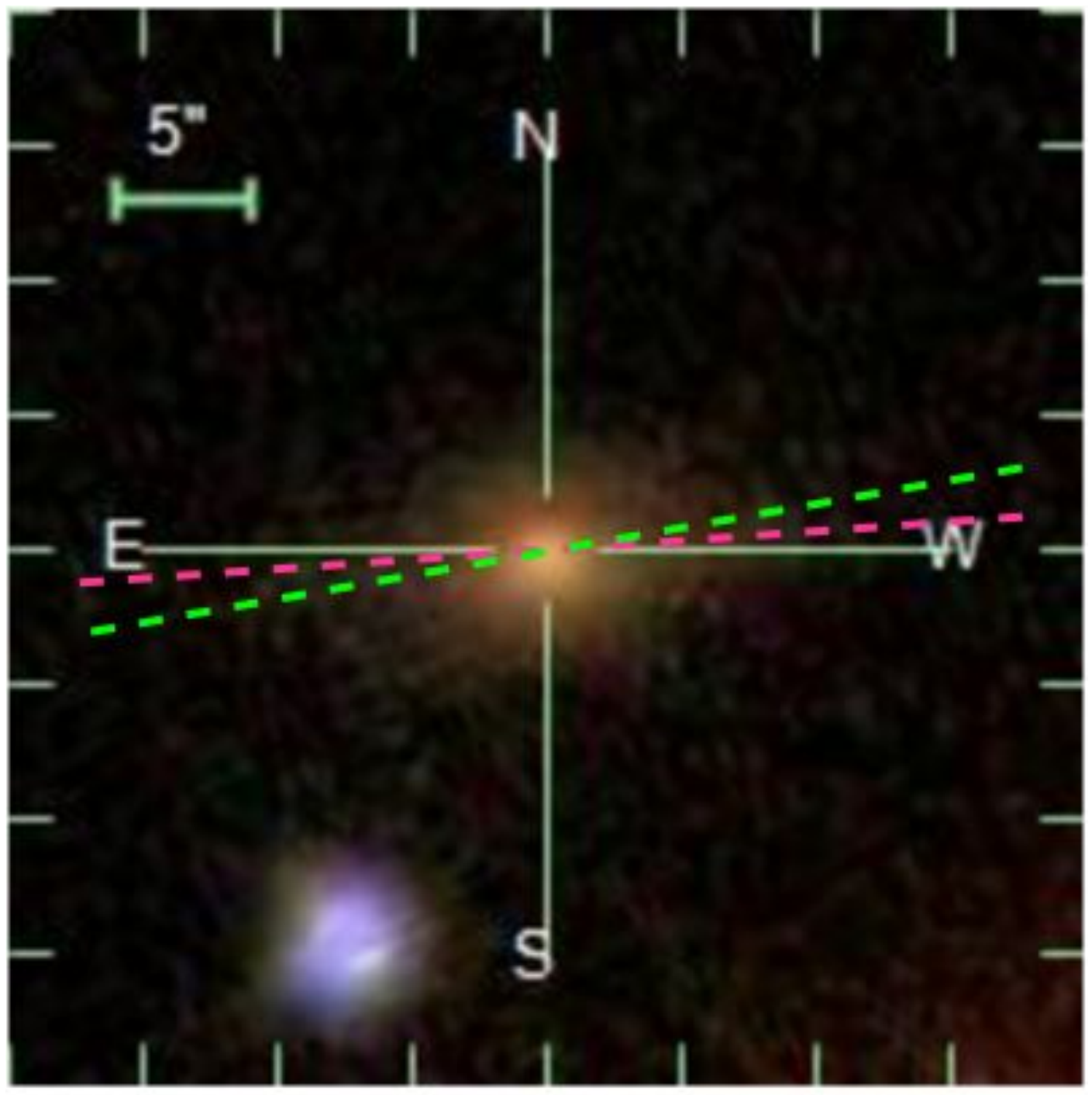}}}$

\caption{The SDSS images of the galaxies classified as `Ambiguous' (J0802+3046, left) and `Rotation Dominated + Obscuration' (J0736+4759, right) in Figure~\ref{proofconcept}. The pink dashed line marks the photometric major axis of the galaxy as measured from SDSS $r$-band photometry. The green dashed line marks the position angle of the [OIII] emission as measured in this work. The [OIII] emission for J0802+3046 is not aligned with the photometric major axis of the galaxy, so we classify it as `Ambiguous' even though in this case all other characteristics of the emission profiles are consistent with a rotation-dominated NLR. Conversely, the [OIII] emission for J0736+4759 is aligned with the photometric major axis, so we classify it as rotation-dominated. Alignment is an important classification tool in distinguishing between rotation-dominated and ambiguous classifications. \label{Alignment}}
\end{center}

\end{figure*}

For each position angle of each galaxy, we measure radial velocity, velocity dispersion, number of kinematic components, and alignment for each spatial row within the Akaike width (Table~\ref{tabkinnum}, for PA 1 and PA 2). Then, we calculate the number of rows that have $>2$ kinematic components within the spatial center (as measured by the stellar continuum) of the galaxy. We classify a galaxy as having $>2$ kinematic components if more than half of the rows of the Akaike width are best fit with $>2$ components (Table~\ref{tabkin}). We record velocity dispersion for both the one Gaussian fit ($\sigma$) and the individual components of the two Gaussian fit ($\sigma_1$ and $\sigma_2$). In Table~\ref{tabkin} we determine if the dispersion of the single Gaussian fit is less than or greater than 500 km s$^{-1}$. We repeat this for each individual velocity dispersion of the two Gaussian fit. Next, in Table~\ref{tabkinnum} we list the position angle of the galaxy (photometric major axis in SDSS $r$-band) and the position angle of the [OIII]$\lambda 5007$. The photometric major axis measurements are reported in \citet{MS2015}, originating from \citet{Comerford2012}. These position angles each have an error of $\sim 7^{\circ}$ associated with the measurement, thus they are classified as aligned if these position angles are within $3\sigma$ ($20^{\circ}$). After recording the quantitative measurements, we classify the galaxies in Table~\ref{tabkin}. We present a cartoon diagram of the kinematic origin of the [OIII]$\lambda 5007$ double peaks for three of the main classification categories (`Outflow', `Rotation Dominated + Disturbance', and `Rotation Dominated + Obscuration') in Figure~\ref{oiiigals}. We show an example of the profiles of the five main classifications, `Outflow', `Outflow Composite', `Rotation Dominated + Obscuration', `Rotation Dominated + Disturbance', and `Ambiguous' in Figures~\ref{proofconcept1} and~\ref{proofconcept}. We show the SDSS images to demonstrate the power of the alignment classification tool to separate `Ambiguous' from rotation-dominated classifications in Figure~\ref{Alignment}.

We take the error on each measured quantity into consideration. We use a superscript to identify galaxies that fall into a given category but based upon the error bars are on the edge of this classification (within 1$\sigma$) in Table~\ref{tabkin}. We include these galaxies in their respective categories for later statistical analysis in Section~\ref{results kin class}.

This quantitative classification system allows us to use the spatial information from longslit spectra to determine the origin of the disturbed kinematics that produce a double-peaked profile. We track the emission lines across the spatial extent of the galaxy and use the behavior of the ionized gas both at each individual position and across the galaxy as a whole to determine the structure responsible for the integrated double peaks. For the first time, we are able to make a distinction between rotation-dominated spectra and outflow-dominated spectra. Additionally, we are able to separate rotation-dominated and outflow-dominated spectra into further kinematically descriptive categories that explain the origin of the double peaks. After classifying the full sample, we find 26/71 Outflow Composite galaxies, 35/71 Outflow galaxies, 4/71 Rotation Dominated galaxies (1/71 Rotation-Dominated + Obscuration, 3/71 Rotation-Dominated + Disturbance), and 6/71 Ambiguous galaxies in this sample. We comment on the distribution of properties and the success of the classification for our kinematic classification technique in Section~\ref{results kin class}.

\section{Results}
\label{all results}
\subsection{Kinematic Classification}
\label{results kin class}
We are able to successfully classify 92\% of the galaxies as either outflow- or rotation-dominated using our stand-alone longslit classification technique. This technique utilizes the spatially-resolved spectra to probe individual locations in the NLR and successfully identifies the nature of the NLR based upon this spectral information alone. We find from the 71 galaxy sample that 6\% of the galaxies (4/71) are Rotation Dominated (1/71 has an obscuration and 3/71 are disturbed), 49\% (35/71) are Outflows, 37\% (26/71) are Outflow Composites, and 8\% (6/71) are Ambiguous (having some combination of outflow, inflow, or rotation-dominated kinematic components).

\citet{Shen2011} conduct a similar classification with optical slit spectroscopy and near-infrared (NIR) imaging of 31 double-peaked AGNs and find 50\% of their sample are classified as having a single NLR, 10\% are candidate dual AGNs, and the remaining 40\% are ambiguous. \citet{Fu2011} use resolved spectroscopy to show that a single AGN with disturbed `gas kinematics' can produce 70\% of the double-peaked profiles. \citet{MS2015} find that 75\% of the double-peaked profiles are produced by `gas kinematics' (including 70\% outflows and 5\% rotating NLRs), 15\% are dual AGNs, and 10\% are ambiguous. \citet{Blecha2013} find from their hydrodynamic simulations that only a minority of double-peaked NLRs result directly from two distinct NLRs associated with two AGNs orbiting a central potential. Most are associated with complex gas kinematics or rotating gas disks. 

Our results agree with these findings, produce fewer ambiguous cases, and further separate complex gas kinematics into rotation-dominated and outflow-dominated categories. The majority (86\%) of our sample is dominated by outflow signatures and only a small minority of the subsample is dominated by rotation. We conclude that selecting Type 2 AGNs by double peaks in integrated SDSS spectra is the most successful at selecting outflows. 

We note that we are not fully able to confirm dual AGNs using this technique in isolation without radio data. We discuss the placement of a candidate dual AGN in the Rotation Dominated + Disturbance category in Section~\ref{rot dual AGN}.

We discuss the properties of the galaxies classified as outflow-dominated in Section~\ref{prop outflow} and the implications for NLR outflow theory. We analyze the properties of the rotation-dominated galaxies in Section~\ref{rot analysis} and compare to predictions from the literature.

\subsubsection{Kinematic properties of the Outflow and Outflow Composite galaxies}
\label{prop outflow}

We find that the majority (86\%) of the uniform sample of double-peaked NLR galaxies are dominated by outflows. Double-peaked emission lines in SDSS are far more successful at selecting AGN outflows than two kinematically distinct rotating NLRs associated with a dual AGN. Here we discuss the kinematic properties of Outflows and Outflow Composite galaxies and we compare these findings to the current theory of the structure of NLR outflows.

\begin{figure*}
\begin{center}
\includegraphics[scale=0.2]{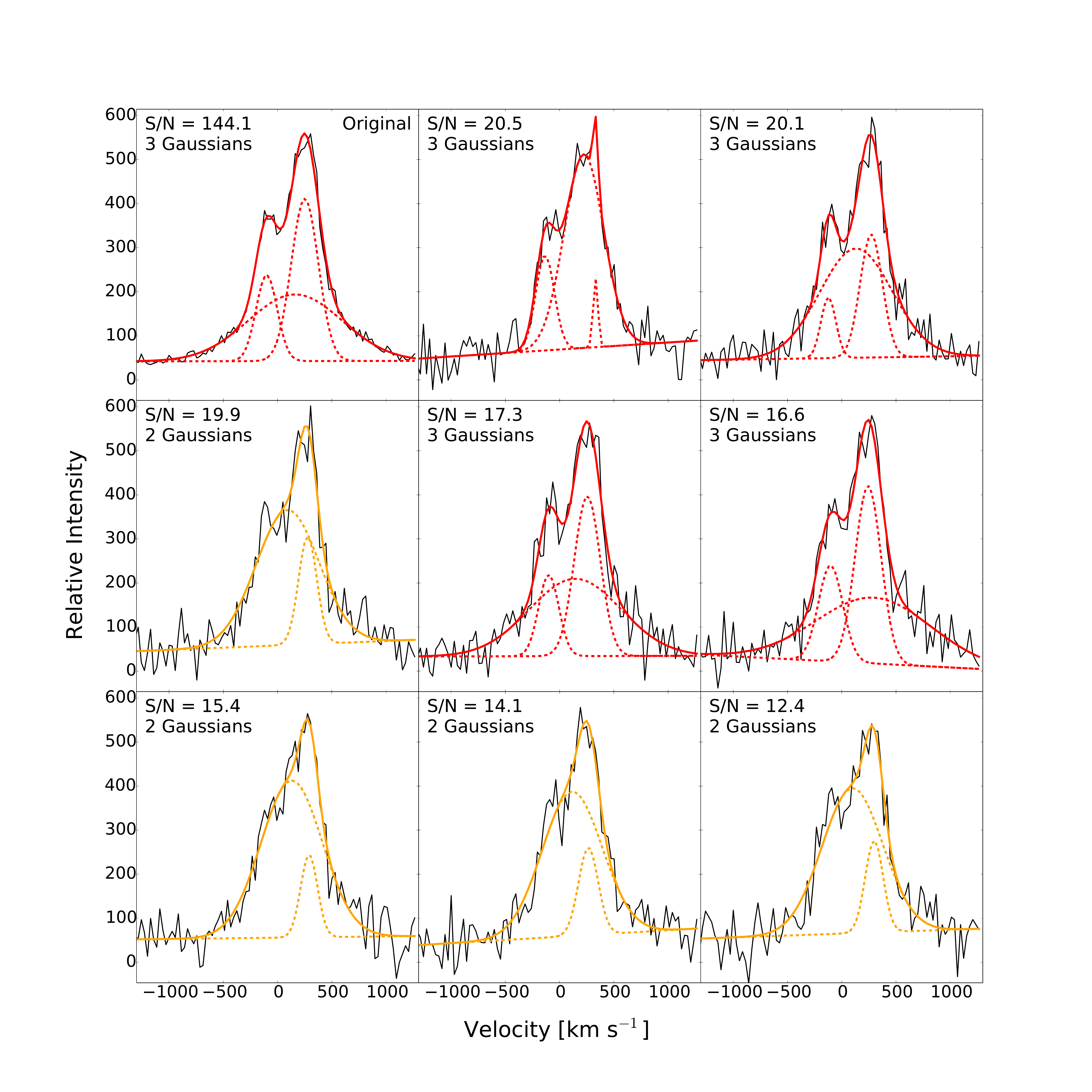}
	
\caption{An example of introducing simulated Gaussian noise into a spatial row of the spectrum of J0803+3926. The original spectrum (upper left) is best fit by three Gaussians. All other panels list the S/N after adding different realizations of simulated and the number of Gaussians that give the optimal fit. If three Gaussian components are the best fit, the fit is shown in red, where the solid red line is the integrated fit and the dashed red lines are the individual component fits. If two Gaussian components provide the best fit, the fit is shown in orange. This galaxy is classified as Outflow Composite. However, when Gaussian noise is introduced into the spectrum, decreasing the S/N, we find that the spectrum is sometimes best fit by two components. Thus, S/N affects the classification of Outflow or Outflow Composite. Note that the effect shown here is not linear; there is no S/N threshold above which a given galaxy will always be classified as Outflow Composite. \label{signalexperiment}}
\end{center}

\end{figure*}

Roughly half of the outflow-dominated galaxies are classified as Outflow. We add the caveat that the classification of Outflow or Outflow Composite is sensitive to S/N; for galaxies with lower S/N, we often do not have enough flux to distinguish the low flux wings of the profiles from the background. We therefore find that these lower S/N galaxies are often better fit with two Gaussians than three. To confirm that the number of Gaussian components is sensitive to S/N, we perform an experiment. We introduce Gaussian noise at a level equivalent to 10\% of the flux to spectra that are best fit with a 3 Gaussian component fit. We complete 100 iterations of this random noise introduction and find that 37\% of the time we recover a three component fit. The rest of the time we find that a 2 Gaussian component fit is better. Figure~\ref{signalexperiment} shows an example simulation of decreased S/N and its effects on the number of Gaussian components that provide the optimal fit for a spatial row. 

We note that although using the number of Gaussian components as a classification tool is sensitive to S/N, we still find it useful as a method of constraining the number of emission knots observed in an outflow. Although S/N can influence this classification, galaxies may intrinsically have only two kinematic components and be true Outflows. However, due to the S/N sensitivity, in some cases the number of components fit is a lower limit, so the galaxies that are classified as Outflow could be classified as Outflow Composites if they were observed with longer integration times. Due to these considerations, we choose to analyze Outflow Composites and Outflows together and refer to this combined category as outflow-dominated in our discussion. 

\begin{deluxetable*}{llllll}
\tabletypesize{\scriptsize}
\tablewidth{0pt}
\tablecolumns{6}
\tablecaption{Kinematic Classification Statistics}
\tablehead{
\colhead{Properties}&
\colhead{Rotation} &
\colhead{Ambiguous}& 
\colhead{Outflow} & 
\colhead{Outflow }\\
 	& Dominated &  & & Composite }
\startdata

	$>2$ Gaussian Components & 0/4 (0\%) & 2/6 (33.3\%) &0/35 (0\%) & 26/26 (100\%) \\

	$V_r > 400$ km s$^{-1}$& 0/4 (0\%) & 0/6 (0\%) &9/35 (25.7\%) & 2/26 (7.7\%) \\

	$\sigma > 500$ km s$^{-1}$& 3/4 (75\%) & 6/6 (100\%) & 28/35 (80.0\%) & 19/26 (73.1\%)\\

	$\sigma_1$ or $\sigma_2 > 500$ km s$^{-1}$& 0/4 (0\%) & 0/6 (0\%)  & 35/35 (100\%) & 26/26 (100\%)\\

	Aligned? (Disk and [OIII])& 4/4 (100\%) & 0/6 (0\%) &17/35 (48.6\%) &10/26 (38.5\%)\\

\enddata
\tablecomments{The statistics of the properties for each kinematic classification. Row 1: The fraction of galaxies in each classification that are best fit by more than 2 Gaussian components. Row 2: The fraction of galaxies with a radial velocity of the single Gaussian component fit in excess of 400 km s$^{-1}$. Row 3: The fraction of galaxies in each classification that have a single Gaussian component velocity dispersion greater than 500 km s$^{-1}$. Although this property is not used in the classification process, we discuss its value for the different classification categories below.  Row 4: Velocity dispersion, but for the individual components of the two Gaussian fits. Row 5: The fraction of galaxies where the position angle of the [OIII]$\lambda 5007$ emission is aligned (within 20$^{\circ}$) with the photometric major axis of the galaxy. }
\label{table:props}

\end{deluxetable*}

We present the kinematic properties of all of the classification categories in Table~\ref{table:props}. We focus here on the properties of the outflow-dominated galaxies. We find that 18\% of the outflow-dominated galaxies have a single Gaussian radial velocity in excess of 400 km s$^{-1}$, 77\% have an overall velocity dispersion greater than 500 km s$^{-1}$, and 100\% have a single component with a velocity dispersion in excess of 500 km s$^{-1}$. Some profiles show the discrepancy where a single Gaussian velocity dispersion is less than the two individual velocity dispersions for the two Gaussian fit. This is because these are double-peaked profiles that are often separated significantly in velocity space. When we fit a single Gaussian component, it sometimes encompasses only one of the two components, while the two Gaussian fit is more sensitive to underlying high velocity dispersion low flux wings. 

 We find that velocity dispersion of the individual components of the two Gaussian fit, $\sigma_1$ or $\sigma_2$, is the most powerful tool in identifying outflow-dominated galaxies. In a few galaxies, radial velocity is also important in identifying knots of emission moving at velocities in excess of a rotation-dominated NLR. However, all of the outflow-dominated galaxies that have a radial velocity in excess of 400 km s$^{-1}$ also have individual components of the two Gaussian fit with velocity dispersions in excess of 500 km s$^{-1}$. We determine that $\sigma_1$ and $\sigma_2$ are more useful as a probe of the bulk motion of the outflow than V$_r$. The knots of emission that achieve velocities in excess of 400 km s$^{-1}$ could be the faintest components of the outflow, and remain unidentified in our Gaussian fitting and classification method for galaxies with lower S/N. Velocity dispersion is a more consistent identifier of outflow-dominated galaxies because it describes the bulk properties of the walls of the outflow as opposed to being related to extremely low surface brightness knots of fast-moving gas.

We found that although we did not directly use the velocity dispersion of a single Gaussian fit ($\sigma$) in our classification, 77\% of the outflow-dominated galaxies had overall velocity dispersions in excess of 500 km s$^{-1}$. However, 75\% of the rotation-dominated galaxies also had this property. This is unsurprising since rotation-dominated galaxies with a disturbance should have an overall velocity dispersion that exceeds the value for an undisturbed rotating disk. The single velocity dispersion $\sigma$ was not as useful as $\sigma_1$ and $\sigma_2$ in discriminating between rotation-dominated and outflow-dominated galaxies.

\subsubsection{Properties of the rotation-dominated galaxies}
\label{rot analysis}
We identify four galaxies as rotation-dominated. The origin of the double peaks is disturbed kinematics from a bar, spiral, or a possible dual AGN in 3/4 cases and an obscuration in 1/4 cases. In other words, the category `disturbed' accounts for all kinematic deviations of a rotation-dominated profile from a single rotating disk. We distinguish between an obscured rotation-dominated galaxy and a disturbed rotation-dominated galaxy using the relative nonparametric asymmetry values of the profiles of each galaxy. Although we measure the relative asymmetry parameter for all galaxies, we only use it as a classification tool to distinguish between different kinematic origins of rotation-dominated galaxies.

\begin{figure*}
\begin{center}
\includegraphics[scale=0.3]{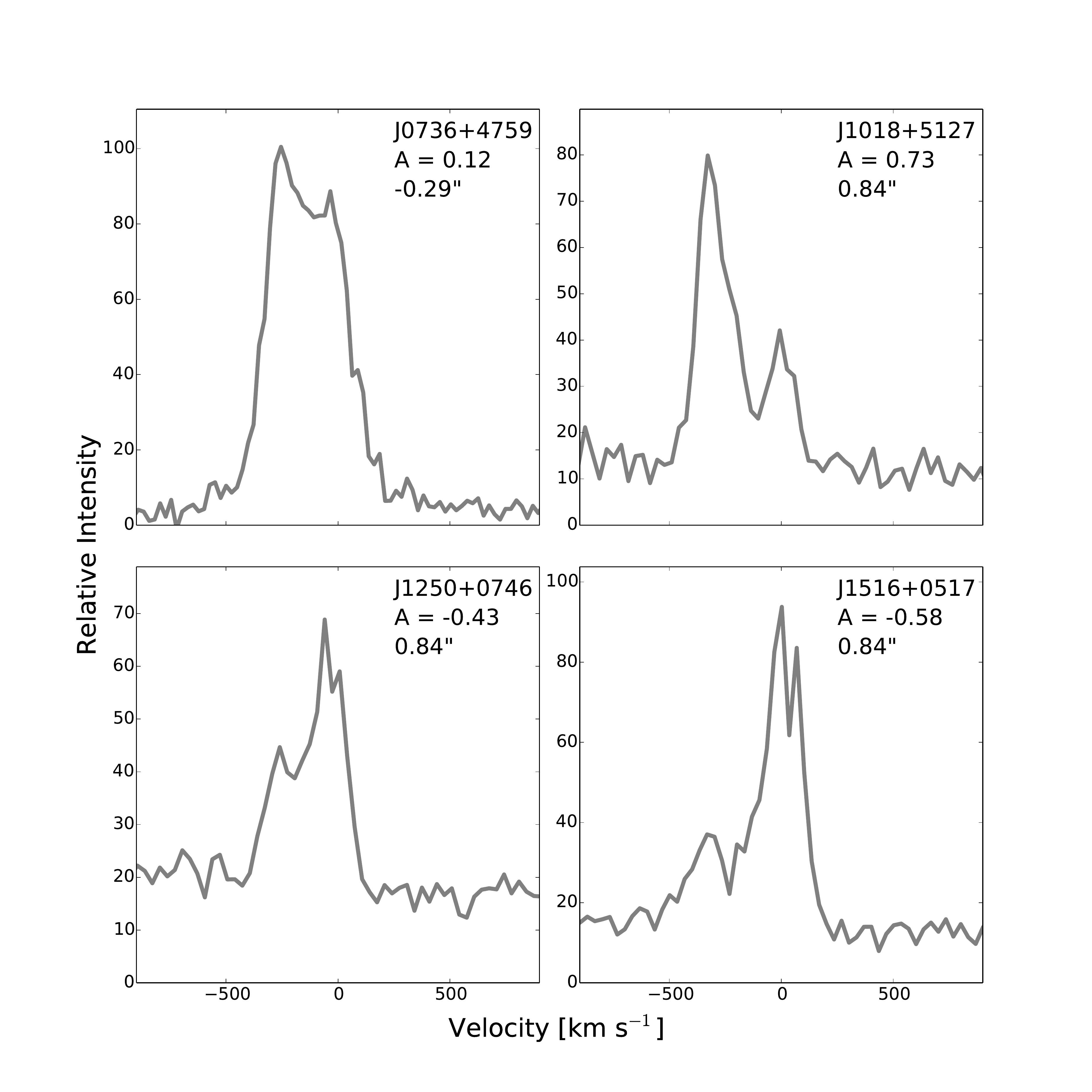}
	
\caption{ The maximum nonparametric asymmetry values (A) and their corresponding spatial row for each of the four rotation-dominated galaxies. We classify the first as symmetric (J0736+4759) and the final three (J1018+5127 , J1250+0746, and J1516+0517) as asymmetric based upon a statistical analysis of the asymmetry values. \label{assy_3} }
\end{center}

\end{figure*}

We use a statistical quantitative method to classify rotation-dominated galaxies as asymmetric or symmetric. We determine the asymmetry values for each galaxy by measuring an asymmetry value for all spatial positions of each position angle. We then take the asymmetry value to be the maximum asymmetry value from either PA. We show this maximum asymmetry value for the four rotation-dominated galaxies in Figure~\ref{assy_3}. The rotation-dominated galaxies have asymmetry values of 0.12 (J0736+4759), 0.57 (J1018+5127), -0.43 (J1250+0746), and -0.72 (J1516+0517). Note that we compare the absolute value of the asymmetry value because a negative or positive value indicates if the profile has blue or red wings, respectively. Using the values of asymmetry for the rotation-dominated galaxies, we can compare the population mean of the entire sample of 71 galaxies (0.43). Given the sample standard deviation (0.15) for this measurement, we can initially conclude that J0736+4759 is about two standard deviations below the mean and the remaining three rotation-dominated galaxies are consistent with the mean (J1250+0746), one standard deviation greater than the mean asymmetry (J1018+5127) and about two standard deviations above the mean (J1516+0517).

We build upon these initial conclusions using the Analysis of Means (ANOM) statistical test. ANOM allows us to set a confidence interval and test for differences in means between subsamples. We plan to group J1018+5127, J1250+0746, and J1516+0517 together and prove that the mean asymmetry of this group is statistically different than the asymmetry of the galaxy J0736+4759, which we have preliminarily demonstrated has a relatively symmetric profile when compared to the other rotation-dominated galaxies. We define the confidence interval and population mean used in ANOM as:

$$ \bar{X} \pm h_{c,n_j} \sqrt{\frac{\sigma_p^2 ( c-1)}{n}}$$
where $\bar{X}$ is the mean of the full sample (71), $c$ is the number of groups between which we wish to compare means (2), $n_j$ is the sample size for group $j$, $h_{c,n_j}$ is the critical value for Nelson's $h$ statistic with $c$ groups and $n_j$ observations per group, $\sigma_p^2$ is the pooled variance of the overall sample, and $n$ is the total number of observations. Note that $n_j$ is set by the smallest sample size of the groups (1). The smallest group sample size is set by J0736+4759 with one measurement. Thus, in our case, $h_{2,1} =12.7$ for a 95\% confidence value.

We obtain a 95\% confidence interval around the sample mean of $[0.19, 0.66]$. We find that the A value of J0736+4759 (0.12) is significantly less than the mean of the overall sample and that the mean of the A values for the group including J1018+5127, J1250+0746, and J1516+0517 (0.58) is consistent with mean of the overall sample. J1018+5127, J1250+0746, and J1516+0517 are classified as asymmetric rotation-dominated galaxies quantitatively according to sample statistics. We classify them as Rotation Dominated + Disturbance. We classify J0736+4759 as Rotation Dominated + Obscuration.

\citet{Liu2013} use an identical definition of nonparametric asymmetry and find values of asymmetry with a maximum around 0.4 for their energetic outflows. Therefore, our quantitative asymmetry cut is even higher, meaning that our rotation-dominated galaxies with a disturbance are even more asymmetric when compared to other work. 

We can also confirm that the three galaxies that we classify as asymmetric under our nonparametric measurement would be classified as asymmetric by \citet{Smith2012} using an alternate definition of asymmetry. \citet{Smith2012} use a ratio of the red and blue flux components of their double-peaked profile fits to quantify asymmetry for their sample of ``equal-peaked'' AGNs (EPAGNs). Specifically, they classify a symmetric EPAGN as one with a value of $0.75 \leq F_r/F_b \leq 1.25$, where $F_r$ is the flux of the redder Gaussian component and $F_b$ is the flux of the bluer component. Their quantitative classification involves fitting two Gaussians. We prefer the nonparametric method of asymmetry quantification because it accounts for features such as low flux wings in complicated profiles that are best fit by more than two Gaussians. 

\citet{Smith2012} conclude that symmetric EPAGNs from the double-peaked sample are most likely to originate from a single rotation-dominated NLR. Our clearest example of this type of a rotation-dominated galaxy is the galaxy J0736+4759, which is rotation-dominated with a central obscuration due to its high degree of symmetry. The peaks are of equal flux for both position angles across the entire slit. This galaxy is consistent with the EPAGNs discussed in (\citealt{Smith2012}). However, the results of our classification reveal that rotation-dominated objects with double-peaked profiles are more likely to be asymmetric due to disturbances such as spirals, bars, or dual AGNs. Three of our four rotation-dominated galaxies have asymmetric peaks.

We also investigate the H$\alpha$ kinematics of the four rotation-dominated galaxies. We refrain from investigating the H$\alpha $ kinematics of all the galaxies in this sample because it is beyond the scope of this work. However, it is a useful exercise to compare the H$\alpha$ kinematics to the [OIII]$\lambda5007$ kinematics for the rotation-dominated profiles because ionized gas dominated by rotation should exhibit gas kinematics that are identical to the stellar kinematics of the stars in the disk. Note that although H$\alpha$ traces both the NLR and stellar kinematics, in the case of a rotation-dominated NLR both the stellar kinematics and gas ionized by the AGN should be consistent with rotation. In other words, if the ionized gas is coincident with the stellar disk, the H$\alpha$ and [OIII]$\lambda 5007$ kinematics should be identical.

\begin{deluxetable*}{lllll}[h]
\tabletypesize{\scriptsize}
\tablewidth{0pt}
\tablecolumns{5}
\tablecaption{$\Delta$V for [OIII]$\lambda 5007$ and H$\alpha$}
\tablehead{
\colhead{SDSS ID} & 
\colhead{[OIII]$\lambda 5007$}  &
\colhead{[OIII]$\lambda 5007$} & 
\colhead{ H$\alpha$}  &
\colhead{H$\alpha$}
 \\
&	$\Delta$V [km s$^{-1}$]  &$\Delta$V [km s$^{-1}$] &$\Delta$V [km s$^{-1}$] & $\Delta$V [km s$^{-1}$] \\
	& PA 1 & PA 2 & PA 1 & PA 2 }

\startdata

J0736+4759$^a$ & 274 $\pm$ 6 &283 $\pm$ 6 &289 $\pm$ 118&273  $\pm$190\\

J1018+5127  &298 $\pm$ 4 &285 $\pm$ 4 &312 $\pm$ 79   & 312 $\pm$ 79\\

J1250+0746  &211 $\pm$ 8 & 218 $\pm$ 4  &  184 $\pm$ 2  & 158 $\pm$ 58\\

J1516+0517 &292 $\pm$ 30 & 285  $\pm$ 50 &320 $\pm$ 81 & 309 $\pm$ 61 \\

\enddata

\tablecomments{Table of the separation in velocity space ($\Delta$V) between the blue and red Gaussian component fits for the four rotation-dominated galaxies. Column 1: Galaxy name. Column 2 and 3: The velocity separation in [OIII]$\lambda 5007$ in km s$^{-1}$ for PA 1 and PA 2 for each galaxy. Column 4 and 5: The velocity separation in H$\alpha$ in km s$^{-1}$ for PA 1 and PA 2.}
\tablenotetext{a}{Due to a restricted wavelength coverage from the observations of J0736+4759, we lack a H$\alpha$ profile and instead compare to the H$\beta$ profile.}

\label{del vs}
\end{deluxetable*}

\begin{figure*}
\begin{center}
	
\includegraphics[scale=0.4]{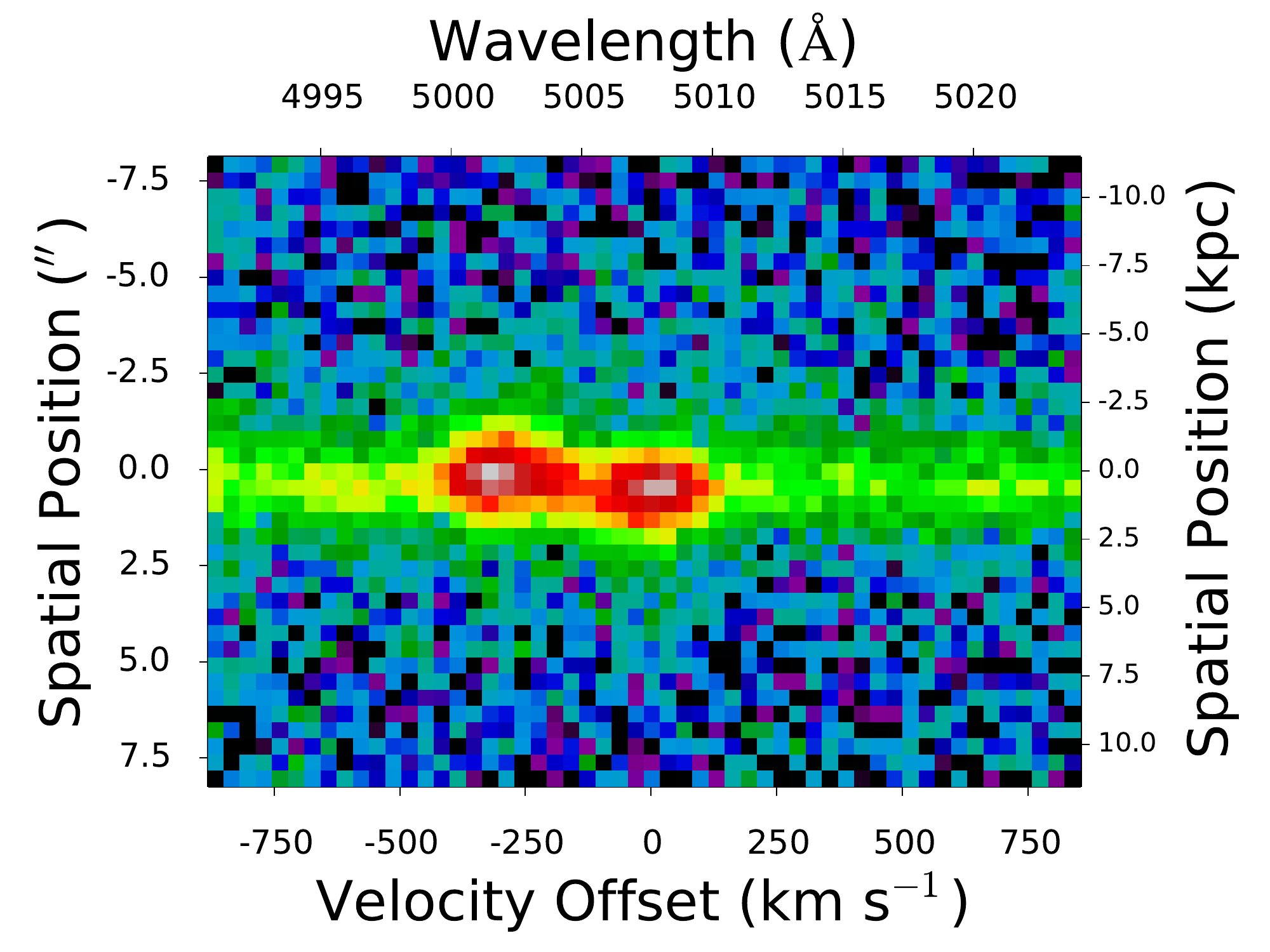}
\includegraphics[scale=0.4]{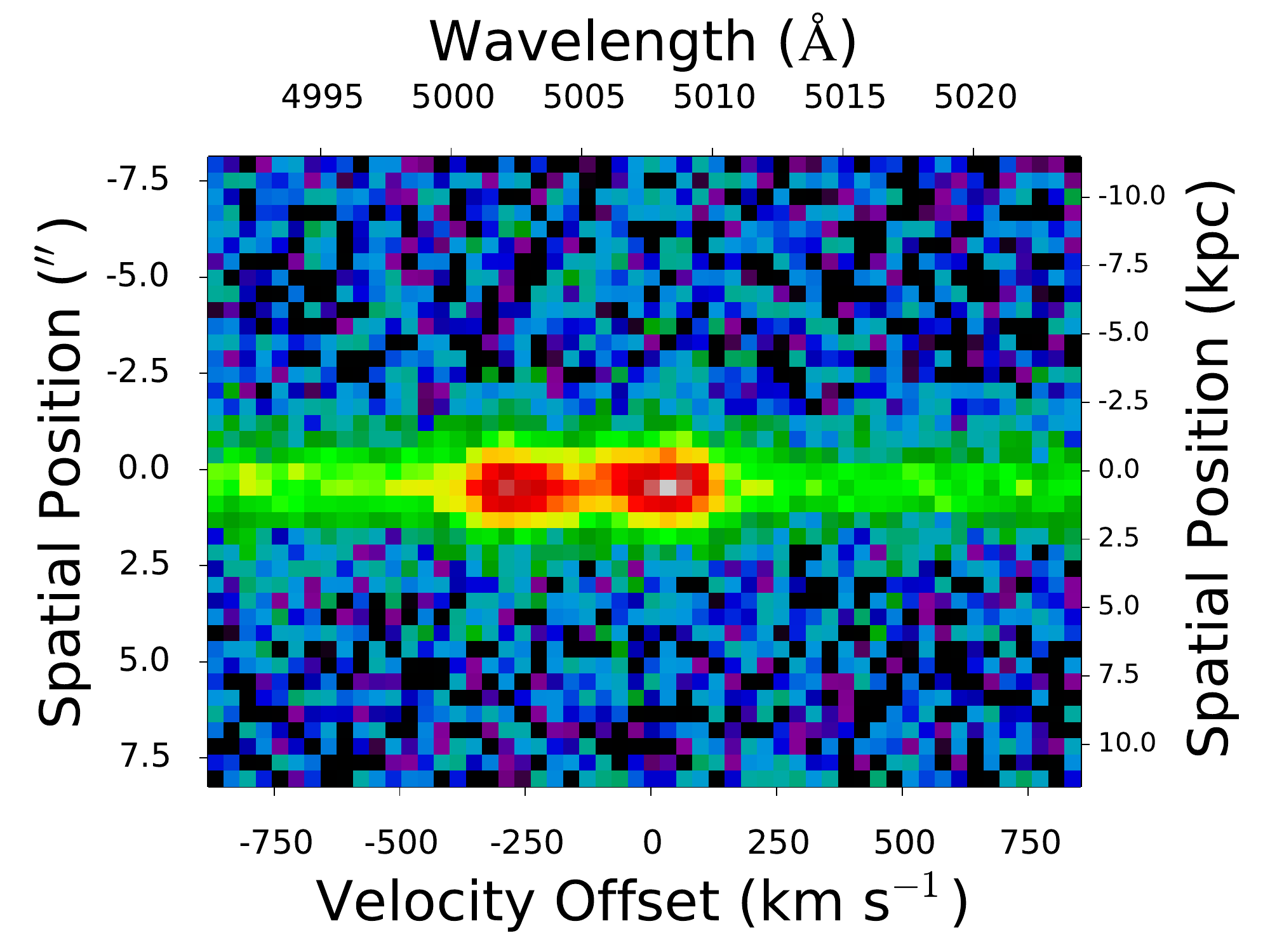}

\includegraphics[scale=0.4]{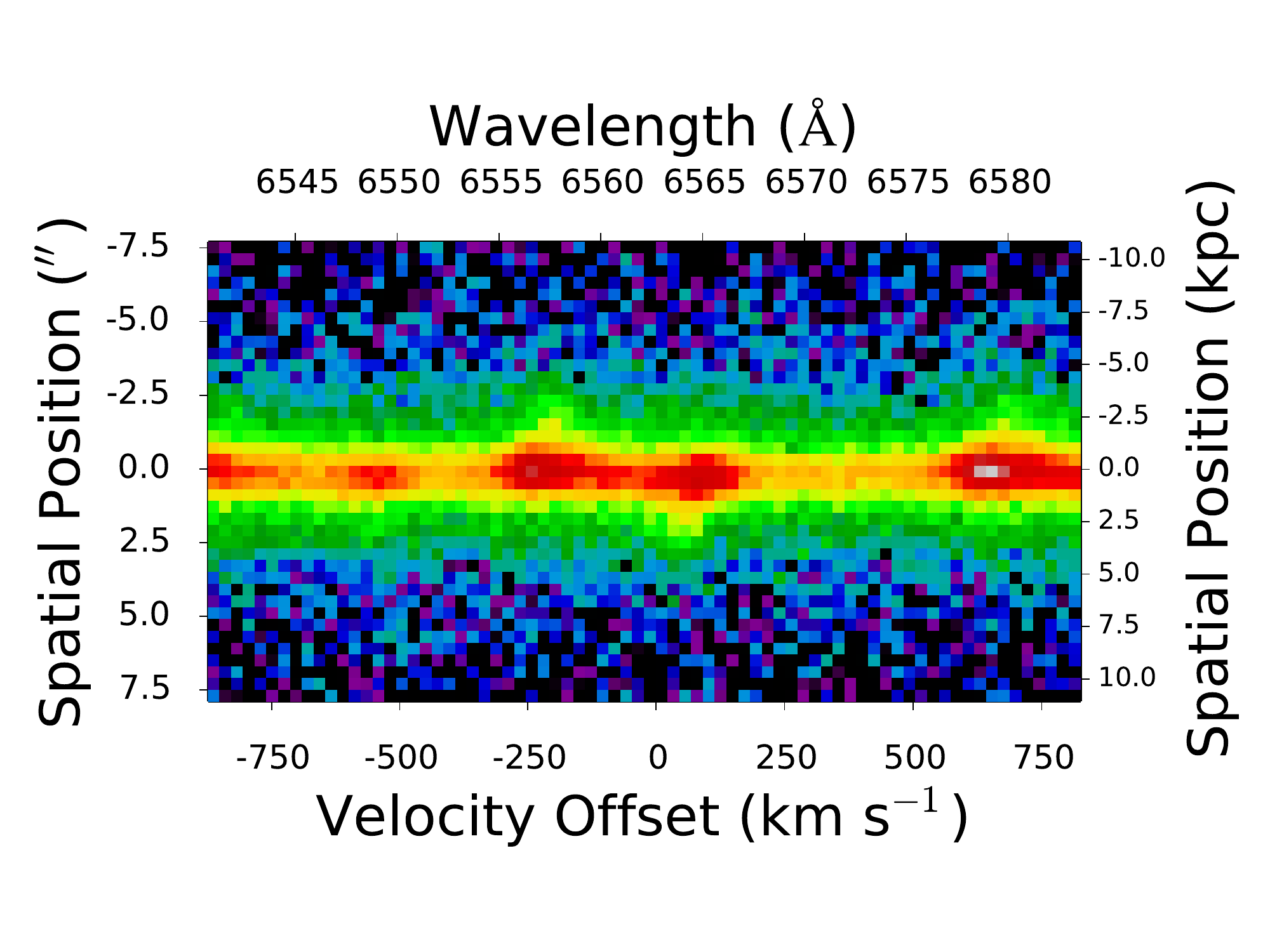}
\includegraphics[scale=0.4]{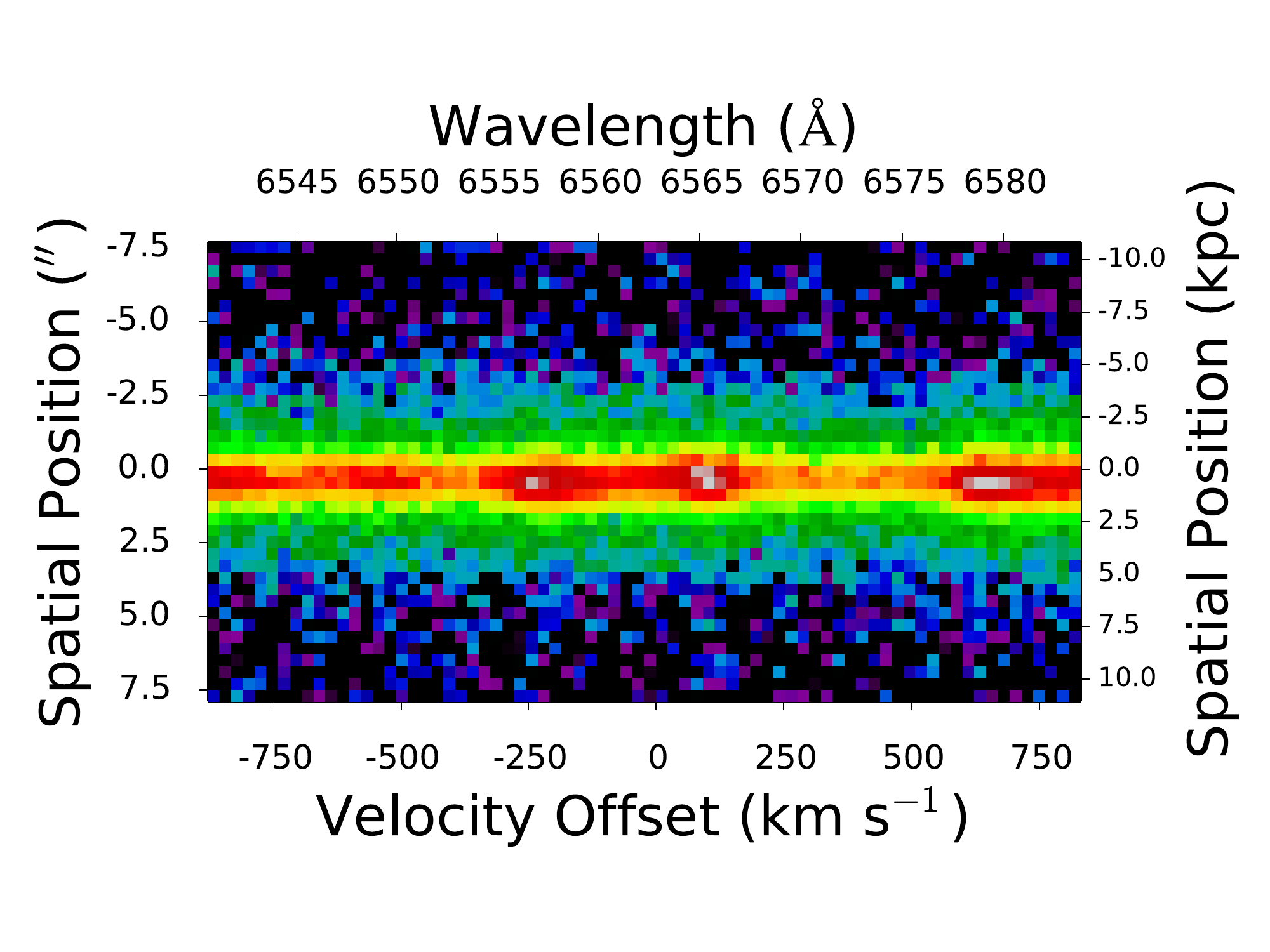}
\caption{The two dimensional longslit spectra for J1018+5127  for each PA (PA 1, left, PA 2, right). The top row shows the [OIII]$\lambda 5007$ profiles and the bottom row shows the H$\alpha$ profiles. The dispersion direction is the x axis in all spectra, with both a velocity axis in km s$^{-1}$ offset from systemic (bottom) and rest frame wavelength axis in \AA (top). The y axis is the spatial direction along the slit reported in both arcseconds (left) and kiloparsecs (right) with the spatial center defined as the center of the galaxy continuum. Note that the line profiles of H$\alpha$ and [OIII]$\lambda 5007$ are similar and consistent in orientation, velocity offset, and velocity dispersion with a gas disk that is coincident with a rotating stellar disk. \label{h alpha J1018} }
\end{center}

\end{figure*}

We find for the four rotation-dominated galaxies that the H$\alpha$ emission is aligned with the [OIII]$\lambda 5007$ emission and that the velocity offsets, velocity dispersions, and relative asymmetry values of the profiles are consistent within the errors (to 3$\sigma$). We present velocity separations ($\Delta$V) between the red and blue Gaussian components of each profile in Table~\ref{del vs} and demonstrate the consistency of the profiles visually in Figure~\ref{h alpha J1018} for J1018+5127. The results verify that the kinematics of these galaxies are indeed dominated by rotation.

\section{Discussion}
\label{all discussion}

\subsection{What are the expected kinematic classifications of dual AGNs?}
\label{rot dual AGN}

This kinematic longslit technique was originally conceived as a method for positively identifying candidate dual AGNs. We cannot exclude dual AGNs from any of the classification categories with this work, but we can offer insight into which categories they could fall. 

A compelling candidate dual AGN could be classified as Outflow or Outflow Composite. Both NLRs will only be visible during the latest kpc and sub-kpc separation stages of dual AGN evolution while both AGNs are accreting simultaneously (\citealt{Blecha2013,vanwassenhove2012,Steinborn2016}). Since observable dual AGNs are by definition in an actively accreting phase of evolution, they can drive outflows and be classified as outflow-dominated. Observationally, past work has found that the optical spectra of confirmed dual AGNs have signatures of disturbed kinematics (outflows and shocks, e.g., \citealt{Mazzarella2012,Engel2010}). Even if one or both distinct NLRs can be described as rotation-dominated, the two AGNs might have unequal luminosities. We would observe this type of galaxy as one narrow component with an associated outflow that dominates a smaller flux second narrow rotation-dominated component from the second AGN. This case of a fainter NLR would be categorized as Outflow Composite or Outflow based upon the number of kinematic components in the outflow and the brightness of the dimmer AGN. 

\citet{Blecha2013} demonstrate that either rotation-dominated NLR of the pair of AGNs could be disturbed by the secondary AGN and appear rotation-dominated with a disturbance. In fact, dual AGNs are more likely to appear observationally as NLRs characterized by disturbed kinematics rather than kinematics dominated by the motion of two SMBHs. We would classify this type of profile as Rotation Dominated + Disturbance. Returning to the picture of the toy model of a dual AGN with two rotation-dominated NLRs, recall that even a theoretical dual AGN with no disturbed kinematics would be classified as Rotation Dominated + Disturbance (e.g., J1018+5127). This theoretical dual AGN would have NLRs that alternate in flux at the position angle of maximum spatial separation (Figure~\ref{dual AGN}). This causes the profile of this type of dual AGN to appear asymmetric for at least one position angle. Lastly, dual AGNs could be classified as Ambiguous because they may demonstrate a lower velocity offset or dispersion, which are properties that are consistent with either an inflow or a less energetic outflow ($\sigma_1$ or $\sigma_2 < 500$ km s$^{-1}$ or V$_r < 400$ km s$^{-1}$).

Even if both rotation-dominated NLRs are visible, \citet{Blecha2013} show that these NLRs would most likely demonstrate a large velocity separation between individual components of order $\Delta V > $ 500 km s$^{-1}$ due to enhanced velocity separation during pericentric passage. These objects would most likely be classified as outflow-dominated by our kinematic classification because the velocity offset between the individual components is so large. This large velocity offset would cause the single Gaussian V$_r$ to be greater than 400 km s$^{-1}$ and this type of galaxy would be classified as an outflow. While the velocity offset limits were derived with the purpose of identifying fast moving outflow components, narrower dual components could also fall into this category. At present, we have no galaxies that fall into this category of being classified as an Outflow due to the velocity offset of two narrow lines with a large velocity separation.

The only category that is unlikely to include dual AGNs is Rotation Dominated + Obscuration. The high degree of symmetry required for this classification is statistically unlikely to be associated with two NLRs producing an outflow or either rotating NLR being disrupted by a secondary AGN. \citet{Blecha2013} predict that equal flux symmetric profiles are most likely associated with an obscured rotating disk, and \citet{Smith2012} support this prediction observationally.

\begin{figure*}
\begin{center}
\includegraphics[scale=0.21]{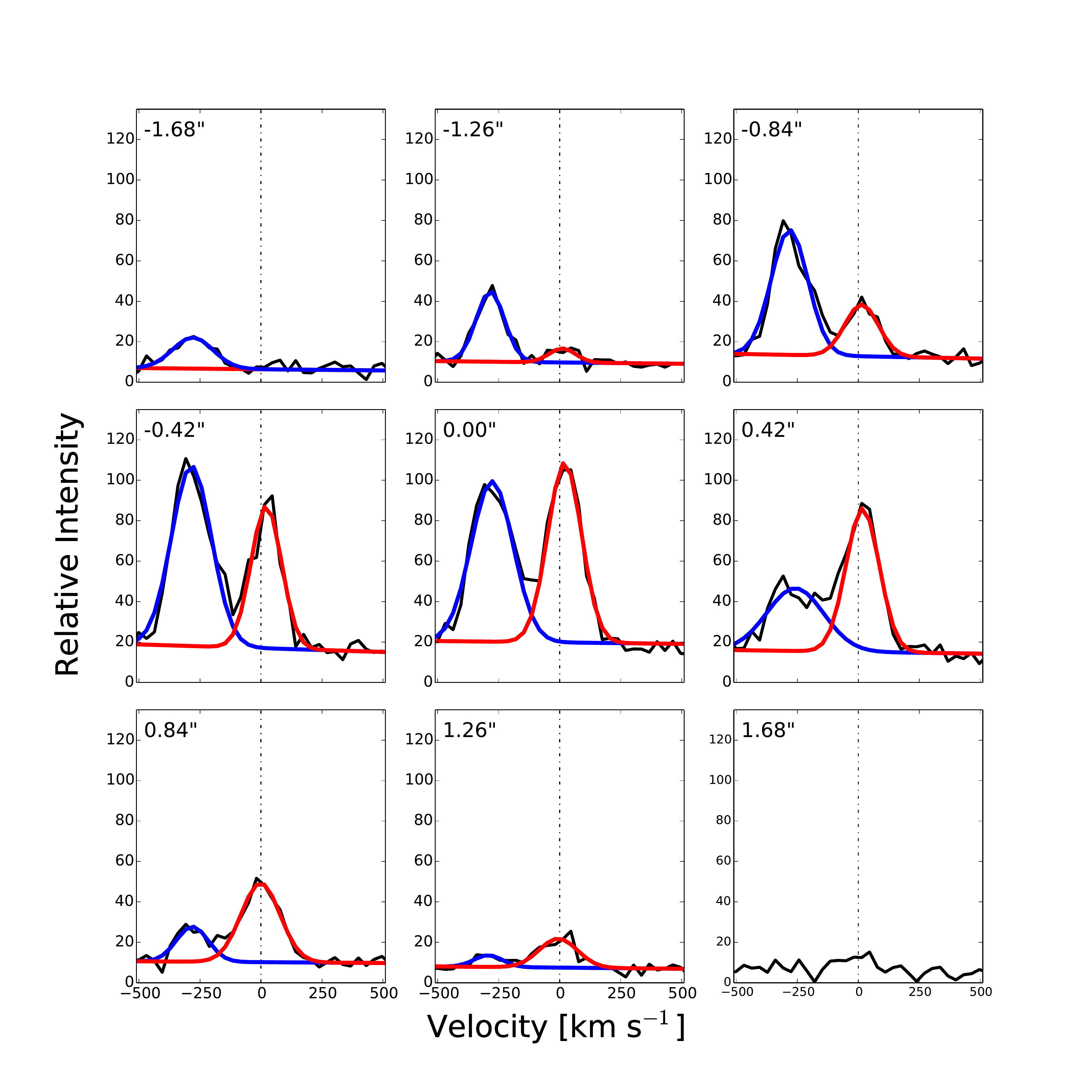}
\includegraphics[scale=0.21]{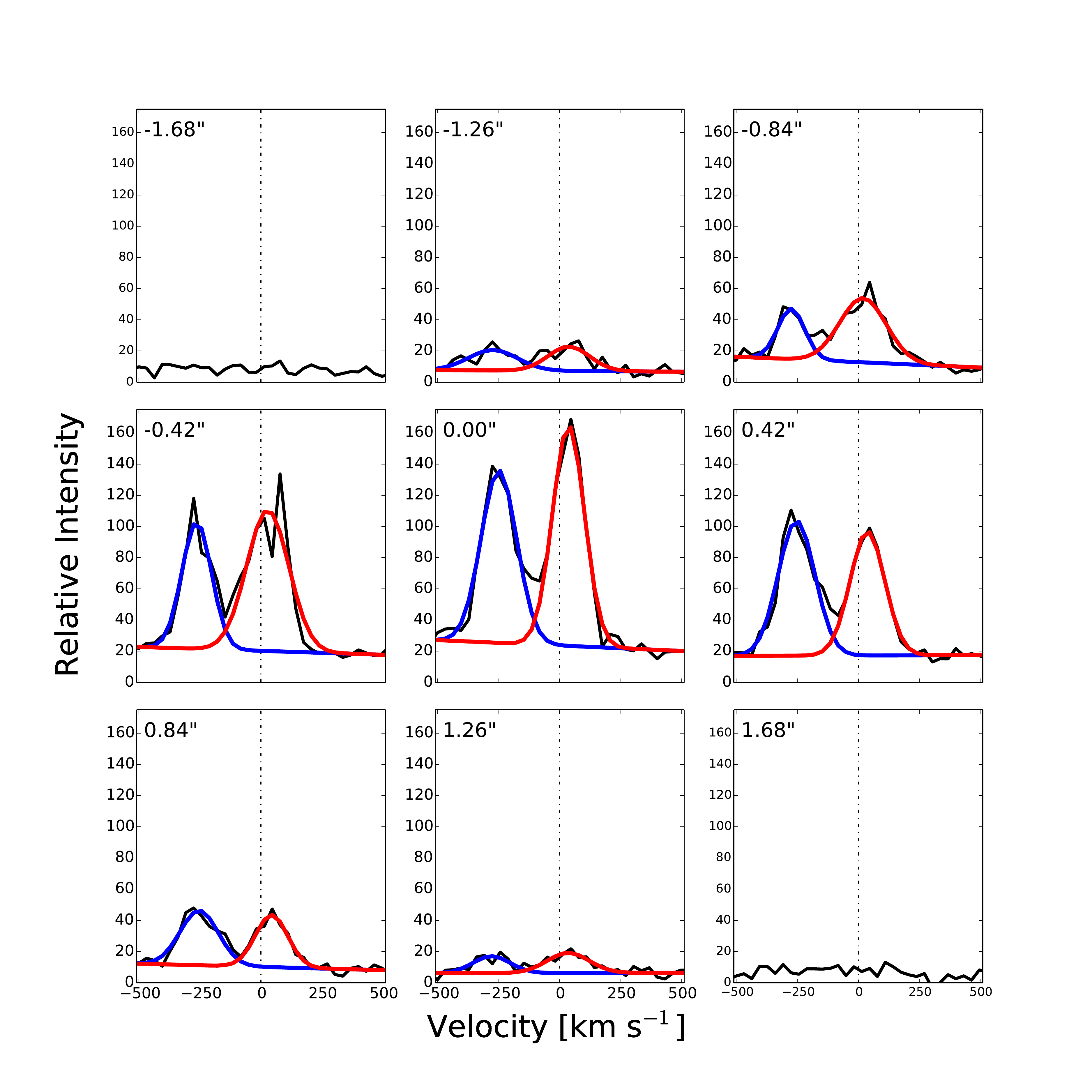}
\caption{The spectral profiles of J1018+5127 at both PAs. We show the PA observed at 22$^{\circ}$ (left) and the PA observed at 112$^{\circ}$ (right). The spatial center of the galaxy (0.0$^{\prime \prime}$) is the central panel, and the spatial position is indicated in the upper right of each panel. We plot the data in black and the two Gaussian fit in blue and red lines for the blueshifted and redshifted Gaussian components of the fit, respectively. Although this galaxy is not confirmed as a dual AGN in the radio data, we demonstrate that the NLRs appear to alternate in flux (left) as expected for the most spatially extended PA of a rotation-dominated dual AGN. On the right, the NLRs could be spatially coincident rotation-dominated NLRs. This galaxy is kinematically classified as Rotation Dominated + Disturbance. \label{J1018 both}}

\end{center}
\end{figure*}

None of the galaxies in this sample are yet confirmed as radio-detected dual AGN, but one galaxy presents a compelling longslit spectrum where each NLR appears to be rotation-dominated. This galaxy, J1018+5127, is undetected in the VLA radio data, so we can place an upper limit on the radio luminosity but cannot reject the presence of a double radio core (M{\"u}ller-S{\'a}nchez et al. in prep.). The longslit [OIII]$\lambda 5007$ profiles of J1018+5127 (Figure~\ref{J1018 both}) are consistent with two distinct rotation-dominated NLRs, and this galaxy is classified as Rotation Dominated + Disturbance. One of the rotation-dominated NLRs is centered at zero velocity and one has a blueshift $\sim 450$ km s$^{-1}$ at all spatial positions. The flux ratio of the NLRs switches along the first PA and the NLRs are roughly equal in flux at the second PA. This is consistent with a dual NLR system with maximal separation nearest the first PA; we measure PA$_{\mathrm{[OIII]}}$ = 22$^{\circ}$, which is nearly aligned with PA$_{\mathrm{gal}}$ = 27$^{\circ}$. This object is most extended in the plane of the galaxy, and this is consistent with a dual AGN interpretation (\citealt{Comerford2012}). Although we can conclude that the longslit kinematic information is consistent with a dual AGN explanation, we stress that this is not a confirmed dual AGN without the complementary double radio core confirmation.

\subsection{A size-luminosity relationship for the NLR reveals the nature of the photoionized region}

\label{r-l rel}
In the study of the structure of AGNs, the nature of the NLR is still uncertain. In this work, we investigate how the extent of the NLR scales with the luminosity of the central source (R$_{\mathrm{NLR}} \propto \mathrm{L}_{\mathrm{[OIII]}}^{\alpha}$). If we can investigate this relationship over a large range of AGN luminosities (by combining our sample with other studies), we can use the strength of this correlation to constrain the ionization structure (ionization parameter) and density structure of the NLR.

The [OIII] luminosity is an accurate probe of the ionization structure and density of the NLR because it is a collisionally excited transition; the line emissivity of [OIII]$\lambda$5007 is set by the electron density $n_e$ and ionization state. Note that although [OIII]$\lambda$5007 can also be produced by stars, BPT diagnostics suggest an AGN origin for the emission in this sample. We use the observed [OIII] luminosity ($\mathrm{L}_{\mathrm{[OIII]}}$; Section~\ref{size-luminosity}) in this work as a probe of the intrinsic luminosity of the NLR.\footnote{We have measured the size-luminosity relationship for our data for both the extinction corrected luminosity ($\mathrm{L}^c_{\mathrm{[OIII]}}$) and the observed luminosity ($\mathrm{L}_{\mathrm{[OIII]}}$) and find that the slope is identical within errors. Note that since other work uses the observed luminosity, we also use $\mathrm{L}_{\mathrm{[OIII]}}$ of our sample of AGNs in the discussion when we compare the luminosity range of our study to other work. }

A positive slope is expected in the size-luminosity relationship if the AGN's radiation is responsible for the photoionization of the NLR. The slope of the relationship, $\alpha$, is influenced by the density structure of the NLR and the ionization parameter, $U$, which is defined as the ratio of the number density of ionizing photons to the number density of electrons:

$$U = \frac{n_{\gamma}}{n_e} = \frac{1}{4\pi \mathrm{R}_{\mathrm{NLR}}^2 c n_e} \int_{\nu_0}^{\infty} \frac{L_{\nu}}{h \nu} d\nu$$

where $n_{\gamma}$ is the number density of ionizing photons, $ \mathrm{R}_{\mathrm{NLR}}$ is the radius of illumination by a central source (Section~\ref{size-luminosity}), $L_{\nu}$ integrated is the bolometric luminosity of the central source over all frequencies, $\nu$ is frequency, and $\nu_0$ is the ionization edge.

Therefore, the relationship between the estimated integrated ionizing luminosity (L$_{\mathrm{ion}}$, in this work we use $\mathrm{L}_{\mathrm{[OIII]}}$ as a proxy for L$_{\mathrm{ion}}$) and radius can be written:

$$ \mathrm{R}_{\mathrm{NLR}} = K {\mathrm{L}_{\mathrm{[OIII]}}}^{0.5} (U n_e)^{-0.5}$$
where $K = (4 \pi c <h\nu>)^{-0.5}$. Determining the slope of the size-luminosity relationship is a direct investigation of both the ionization structure $U$ and the density structure $n_e$ of the NLR, both of which are poorly determined.

Our sample of 71 moderate-luminosity ($40 < \mathrm{log}\ \mathrm{L}_{\mathrm{[OIII]}}$(erg s$^{-1}$)$< 43$) AGNs represents a unique opportunity to investigate this size-luminosity relationship for a uniform sample of moderate luminosity AGNs with resolved NLRs. Our measurement of R$_{\mathrm{NLR}}$ (Section~\ref{size-luminosity}) is a lower limit since we only have spatial data along two dimensions of the galaxy. If neither position angle is exactly aligned with the position angle of maximal extent of the NLR, we are unable to measure the true extent of the NLR (R$_{\mathrm{NLR}}$; Section~\ref{size-luminosity}). However, we note that other studies of the size-luminosity relationship are also limited by resolution and surface brightness considerations (\citealt{Hainline2013,Schmitt2003,Bennert2002,Fraquelli2000,Liu2013a}). 

\begin{figure*}
\begin{center}

\includegraphics[scale=0.8]{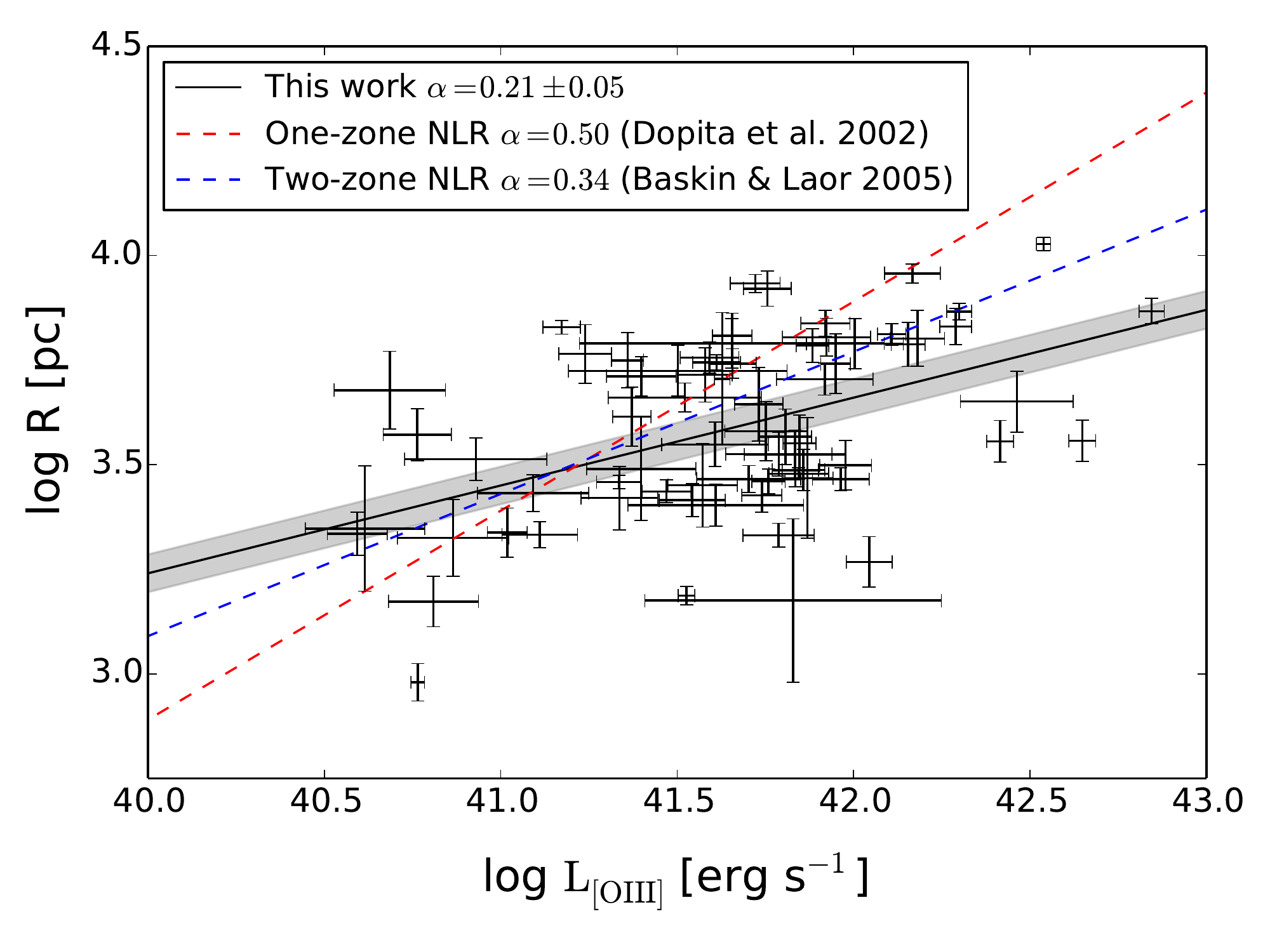}
\caption{Size-luminosity ($\mathrm{L}_{\mathrm{[OIII]}}$) relationship for the NLR for the observed luminosity. We plot the 71 galaxies from this sample in black. We overplot the best fit line in solid black and find a value of $\alpha = 0.21 \pm 0.05$ for the slope of the log-log relationship. We include the confidence interval for this slope in the shaded region. We also plot the predicted slope for a one-zone NLR model in red and a two-zone NLR model in blue. \label{r vs l}}

\end{center}
\end{figure*}

Here we review the slope of the size-luminosity relationship measured by other studies. In a simplistic one-zone model of the NLR, the NLR is described by an isotropic distribution of gas where $n_e$ and $U$ are constant. In this case, the size-luminosity relationship will have a positive correlation and slope of $\alpha = 0.5$ (\citealt{Baskin2005}). \citet{Bennert2002} find a slope of $\alpha = 0.52 \pm 0.06$ for their sample of seven Seyfert 2 galaxies and seven quasars ($41 < \mathrm{log}\ \mathrm{L}_{\mathrm{[OIII]}}$(erg s$^{-1}$)$< 43$) and conclude that a constant density law and a constant value for the ionization parameter describe the NLR. \citet{Hainline2013} predict a limit to the correlation at the higher luminosity extreme of AGNs ($42 < \mathrm{log}\ \mathrm{L}_{\mathrm{[OIII]}}$(erg s$^{-1})< 43$). This is confirmed by \citet{Hainline2014}; they observe quasars at the high luminosity limit of AGNs and observe a flattening in the size-luminosity relationship. They attribute this flattening to a limit in the amount of gas in the gas reservoir of the NLR that is available for ionization. \citet{Liu2013a} confirm this effect and find a flatter slope of $0.25 \pm 0.02$ for a sample of 11 luminous obscured quasars ($42 < \mathrm{log}\ \mathrm{L}_{\mathrm{[OIII]}}$(erg s$^{-1}$)$< 44$). 

However, the flattening at the high luminosity end of the relationship is not the only interesting conclusion from previous work. \citet{Schmitt2003} find a slope of $0.33 \pm 0.04$ for their sample of moderate luminosity ($39 < \mathrm{log}\ \mathrm{L}_{\mathrm{[OIII]}}$(erg s$^{-1}) < 42$) Seyfert 1 and Seyfert 2 galaxies, which is much shallower than the slope of $\alpha = 0.5$ predicted for a constant density law and ionization parameter. They conclude that a single-zone model with a constant density and ionization parameter is not an appropriate representation of the NLR and that a two-zone ionization model for the NLR is a better explanation (\citealt{Baskin2005,Dopita2002}). 
  
In a two-zone ionization model, the NLR can be described by both a matter-bound zone and an ionization-bound zone, where the ionizing photons are pre-processed by passing through a ionization-bound zone. The ionization-bound zone is characterized by a lower ionization parameter and a higher density. The ionization-bound regime is thus optically thick to ionizing radiation. In the most simplistic two-zone models, the ionization-bound zone is confined to a smaller radius and the matter-bound zone exists at spatial positions exterior to this. We will also discuss a more complicated two-zone model where these two zones exist in a clumpier and/or mixed state but note that \citet{Baskin2005} model the radially confined case where denser NLR clouds are confined to a smaller radii and all NLR material is distributed continuously. Most work deals with this simplistic model. \citet{Baskin2005} assume that the majority of emission originates from the outer matter-bound zone where the size-luminosity relationship can be modeled with a slope of $\alpha = 0.34$. Thus, a NLR with mixed matter-bound and ionization-bound zones could be described by a slope between the extremes of $\alpha=0.34$ and $\alpha = 0.5$. Likewise, a NLR with a clumpier distribution of matter-bound material than the continuous assumption would have a slope shallower than $\alpha=0.34$.

Observations support this idea of a changing ionization parameter; the ionization parameter can decrease with radius (e.g., \citealt{Fraquelli2000}), which could be a signature of a matter-bound region at a larger radius of a galaxy. \citet{Liu2013a} confirm that beyond 7 kpc for their quasars sample, the ionization diagnostic [OIII]/H$\beta$ declines. They argue that the NLR is entering a matter-bound regime at these radii, which explains the shallower slope they fit to their size-luminosity relationship ($ 0.25 \pm 0.02$). Consequently, arguing for a two-zone ionization model where an outer matter-bound zone is characterized by a lower density is similar to arguing that the gas reservoir is depleted at these extreme radii.

Shocks may also influence the size-luminosity relationship for AGNs. \citet{MS2015} find a steeper slope of $0.52 \pm 0.14$ for a sample of 18 double-peaked AGNs that includes three confirmed dual AGNs. We note that shocks may steepen this slope by introducing higher-ionization zones. Dual AGNs, which host two interacting NLRs, are most likely to produce shocks that are enhancing this relationship, thus enhancing the [OIII]$\lambda 5007$ emissivity.

From our sample of moderate luminosity AGNs, we confirm a linear correlation with a Pearson correlation coefficient of 0.48 and find a log-log slope of $0.21 \pm 0.05$ (Figure~\ref{r vs l}).\footnote{We also fit this relationship for the intrinsic (corrected) [OIII] luminosity, $\mathrm{L}^c_{\mathrm{[OIII]}}$, and find a slope of 0.24 $\pm $0.05. Our two measurements are consistent within error. We choose to use the observed luminosity value of slope in our discussion to compare with other work (\citealt{Bennert2002,Schmitt2003,Hainline2013,Hainline2014,Liu2013a}). We can reject the null hypothesis that $\alpha = 0$ to a 1\% confidence level. We plot the results of our linear fit in Figure~\ref{r vs l} with the slopes of the one-zone and two-zone ionization models for comparison. We discuss our linear correlation coefficient, fitted slope, and confidence interval on this value both in the context of the one and two-zone ionization models.}

Our results for a fitted slope are inconsistent with the predicted slope of $\alpha =0.5$ for a one-zone ionization model to 3$\sigma$ confidence. Our results cannot be explained by a constant density profile with a constant ionization parameter. Our slope is closer to the predicted slope of 0.34 (within 3$\sigma$ confidence) for the two-zone ionization model of the NLR (\citealt{Baskin2005}). However, our slope is still shallower than $\alpha=0.34$ at a confidence level of 1$\sigma$. We suggest two possible explanations for this discrepancy. 

First, the filling factor of the NLR is not well determined, so if the clouds are sparsely distributed, this would flatten the relationship to values less than the $\alpha=0.34$ slope derived for a simple two-zone ionization model. This simple model assumes a smooth density distribution of the NLR gas and a radial distribution of the two zones. This implies that the simplistic two-zone model may not be adequate to explain our data and we may a require a clumpier or mixed NLR that is not well described by a simplistic radial density profile as in \citet{Baskin2005}.

Alternatively, we measure larger than average NLR extents when compared to NLR size measurements for AGNs with a similar luminosity range ($40 < \mathrm{log}\ \mathrm{L}_{\mathrm{[OIII]}}$(erg s$^{-1}$)$< 43$); see \citet{Hainline2013} and references therein. We find a mean R$_{\mathrm{NLR}}$ of 4.3 kpc, which is comparable to studies of more luminous AGNs (e.g., \citet{Bennert2002} find a mean R$_{\mathrm{NLR}}$ of 4.3 kpc for their $41 < \mathrm{log}\ \mathrm{L}_{\mathrm{[OIII]}}$(erg s$^{-1}$)$< 43$ AGNs and \citet{Hainline2013} find a mean R$_{\mathrm{NLR}}$ of 3.8 kpc for their $42 < \mathrm{log}\ \mathrm{L}_{\mathrm{[OIII]}}$(erg s$^{-1}$)$< 43$ AGNs) and our value of R$_{\mathrm{NLR}}$ is greater than studies of AGNs with a similar luminosity range (e.g., \citet{Schmitt2003} find a maximum R$_{\mathrm{NLR}}$ of 1.6 kpc for their $39 < \mathrm{log}\ \mathrm{L}_{\mathrm{[OIII]}}$(erg s$^{-1}$)$< 42$ AGNs).

Although we refer to our sample as moderate luminosity AGNs, note that some of them have observed luminosities in a range described as higher luminosity ($42 < \mathrm{log}\ \mathrm{L}_{\mathrm{[OIII]}}$(erg s$^{-1}$)$< 43$). We may be probing regions of the NLR where the gas reservoirs available for ionization are limited and the material is better characterized as matter-bound. \citet{Baskin2005} derive the expected slope of 0.34 for a two-zone ionization model for galaxies with a maximum spatial extent of the NLR of 1.3-1.7 kpc and a constant density. Our sample of galaxies may have a relatively larger and/or sparsely populated (less dense) matter-bound region.  These effects could lead to a shallower slope for the relationship. We note that while we cannot fully distinguish between these two scenarios, our results are more consistent with a two-zone rather than a one-zone picture of the NLR. We refer to some combination of a non-constant density or non-radially distributed matter-bound region as a `clumpy two-zone model'.

The positive correlation of this relationship indicates that the AGN itself is the mechanism responsible for ionization of the NLR. The Pearson correlation coefficient for the data (0.48) reflects scatter in the data. This could be a consequence of the lower limit nature of the R$_{\mathrm{NLR}}$ measurement. The error bars show significant uncertainty in both the luminosity measurement and the spatial extent measurement for some galaxies in the sample. Another source of the scatter in our data could be the nature of a two-zone ionizing model for the NLR. While studies mostly describe the matter-bound zone as the outer zone of a galaxy, the matter-bound and ionization-bound zones could exist at different locations in a galaxy, forming a ``clumpier'' picture of an intermixed two-zone NLR. Therefore, individual galaxies in our large sample could have different ionization structures and this could intrinsically produce the scatter.

\section{Conclusions}
\label{conclude}
Based on optical longslit spectroscopy of the complete sample of 71 double-peaked AGNs at $z < 0.1$ in SDSS, we create and implement a classification system for double-peaked NLR emission lines. Our method determines the kinematic origin of the emission at different spatial positions for each galaxy. We present the following conclusions based upon this technique: 
\begin{itemize}

\item Of the sample of 71 galaxies, 6\% have kinematics dominated by rotation, 86\% of the galaxies are dominated by outflows, and 8\% of the galaxies are dominated by some combination of outflows, inflows, and rotation. Our kinematic classification determines that the majority of double-peaked emission lines originate from outflows and succeeds at further determining the properties of the gas outflows and the rotating disks.

\item While we cannot confirm (or exclude) dual AGNs using the kinematic classification method in isolation, we find that dual AGNs can be classified under any category other than rotation-dominated with an obscuration.

\item We find that the 71 AGNs in this sample demonstrate a positive correlation between NLR size and luminosity (R$_{\mathrm{NLR}}\propto \; {\mathrm{L}_{\mathrm{[OIII]}}}^{0.21 \pm 0.05}$). This suggests a two-zone clumpy ionization model for the NLR.

\end{itemize}

The full sample of double-peaked AGNs at $z < 0.1$ have been observed in the radio. Future work will combine the optical and radio data to investigate the orientation of the radio emission and its relationship to the ionized gas. To further investigate the direct effects of feedback on the host galaxies in this sample, we will also introduce an analytic model for the structure of the biconical outflow for the galaxies classified kinematically as outflow-dominated galaxies. Using these models, we will constrain the energetics and momentum entrained in these ionized outflows and discuss their effect on star formation in the host galaxies.

\section{Acknowledgements}

We thank the anonymous referee for a helpful and insightful report that has improved the strength of this work. R.N. is supported by an National Science Foundation (NSF) Graduate Research Fellowship Program (GRFP) Fellowship. M.C.C. was supported by NSF grant AST-1518257. The observations reported here were obtained at the Apache Point Observatory 3.5m telescope, which is owned and operated by the Astrophysical Research Consortium; Lick Observatory, a multi-campus research unit of the University of California; the Hale Telescope, Palomar Observatory as part of a continuing collaboration between the California Institute of Technology, NASA/JPL, and Cornell University; the MMT Observatory, a joint facility of the University of Arizona and the Smithsonian Institution; and the W.M. Keck Observatory, which is operated as a scientific partnership among the California Institute of Technology, the University of California and the National Aeronautics and Space Administration. The W.M. Keck Observatory was made possible by the generous financial support of the W.M. Keck Foundation.

{\it Facilities:} APO (Dual Imaging spectrograph), Keck (DEep Imaging spectrograph), MMT (Blue Channel spectrograph), Lick (Kast spectrograph), Palomar (Double spectrograph)

\bibliographystyle{apj}

\end{document}